\documentclass[%
reprint,
superscriptaddress,
 amsmath,amssymb,
 aps,
 prx,
]{revtex4-2}

\usepackage{graphicx}% Include figure files
\usepackage{dcolumn}% Align table columns on decimal point
\usepackage{bm}% bold math
\usepackage[T1]{fontenc}
\usepackage{placeins}
\usepackage[hidelinks]{hyperref}
\usepackage{url}
\usepackage{xcolor}
\begin{document}

\title{Super Moir\'e Domain Tessellations, Sliding Ferroelectricity, and \\ Reconfigurable Quantum Dot Arrays in Twisted Trilayer Hexagonal Boron Nitride}

\author{Kunihiro Yananose}
 \affiliation{Korea Institute for Advanced Study, Seoul 02455, Korea }

\author{Changwon Park}
\email{Contact author: cwparkphys@ewha.ac.kr}
 \affiliation{ Department of Physics, Ewha Womans University, Seoul 03760, Korea }
 \affiliation{Institute for Multiscale Matter and Systems, Ewha Womans University, Seoul 03760, Korea}

\author{Young-Woo Son}
 \email{Contact author: hand@kias.re.kr}
 \affiliation{Korea Institute for Advanced Study, Seoul 02455, Korea }

%\date{\today}

\begin{abstract}  
At very small twist angles, bilayer moir\'e systems exhibit characteristic stacking domain patterns, where the moir\'e length scale is determined solely by the twist angle. In contrast, the additional stacking and twisting degrees of freedom in twisted trilayer systems give rise to richer and more intricate domain tessellations. In twisted trilayer hexagonal boron nitride (TTBN), the interplay between polar and nonpolar domains and their domain walls is shown to result in unconventional responses to external electric fields, including electric-field tunability of the moir\'e-of-moir\'e or super moir\'e pattern--features absent in bilayer counterparts. We demonstrate that at the vertices of super moir\'e domains, TTBN can support arrays of quantum dots hosting localized quantum harmonic oscillator (QHO) states with diverse spatial symmetries. Futhermore, we show that the shape of the array and the spacing between the localized QHO states  can be dynamically reconfigured by electric fields, enabling facile switching between fully isolated and strongly coupled regimes. The local potentials for the quantum dot state are predicted to be sufficiently deep to support a series of QHO states with nonzero angular momentum. This tunability enables control over the transport of quantum dot states and their interdot coupling, facillitating long-range quantum state transfer. Combined with the feasibility of large-scale fabrication of homogeneous twisted trilayer materials, these properties position TTBN as a promising platform for a wide range of quantum technologies.
\end{abstract}

\maketitle

\section{Introduction}\label{sec:intro}

Substantial progress has been achieved in the development of quantum hardware designed to tackle computationally intractable problems that lie beyond the reach of classical computing architectures~\cite{Ladd2010Nature,Ebadi2021Nature,Bluvstein2022Nature,Bruzewicz2019APR,Monroe2021Nature,Nathalie2021Science,Acharya2025Nature}. 
Multiple platforms have emerged as viable candidates, each with distinct advantages~\cite{Nathalie2021Science}. 
The cornerstone for these implementations lies in precise controls over individual quantum states as well as their mutual interactions~\cite{Nathalie2021Science}. 
This requires the reliable preparation of well-defined quantum objects such as atoms~\cite{Ebadi2021Nature,Bluvstein2022Nature}, ions~\cite{Bruzewicz2019APR,Monroe2021Nature}, superconducting circuits~\cite{Acharya2025Nature}, spin defects~\cite{Pezzagna2021APR} and quantum dots (QDs)~\cite{Philips2022Nature,Burkard2023RMP,Borsoi2024NatNano}. Simultaneously, it also necessitates advanced engineerings for scalable and on-demand deployment of many quantum objects with homogeneous quality controls~\cite{Nathalie2021Science}.

Among these candidates, neutral atom and ion arrays have demonstrated exceptional quantum control capabilities over a large number of individual quantum units~\cite{Ebadi2021Nature,Bluvstein2022Nature,Bruzewicz2019APR,Monroe2021Nature}. In addition,  superconducting circuits have already shown significant advancement~\cite{Acharya2025Nature}. 
Considering well-established nanofabrication technology of semiconductor industry, solid-state implementations, particularly QDs, also hold considerable promise~\cite{Philips2022Nature,Burkard2023RMP,Borsoi2024NatNano}. 
Despite impressive progress in quantum information applications using QD arrays~\cite{Philips2022Nature,Burkard2023RMP,Borsoi2024NatNano}, significant challenges still remain in achieving their large-scale arrangements with uniform electronic properties~\cite{Nathalie2021Science} and their controlled on-demand reconfigurations~\cite{Pan2015PRL,Yan2019AdvPhys} — capabilities that are more readily available in atomic physics platforms~\cite{Bluvstein2022Nature}.

Concurrently, intensive research efforts have focused on engineering quantum states through the controlled stacking of two-dimensional (2D) materials with twist angles between adjacent layers~\cite{Andrei2021NatRevMat,Kennes2021NP}. 
These configurations generate large-scale periodic superlattices, known as moir\'e lattices~\cite{Andrei2021NatRevMat,Kennes2021NP}, that extend beyond conventional atomic periodicity. 
In twisted bilayer systems, these structures exhibit various notable quantum phenomena through interactions between quasiparticles or excitations modulated by moir\'e periodic potentials~\cite{Kennes2021NP,Bistritzer2011PNAS,Cao_correlated_2018Nature}. 
Furthermore, 
the versatility afforded by combining various characteristic 2D materials creates valuable opportunities for realizing unique quantum devices~\cite{Montblanch2023NatNano}.

For quantum phenomena observed in moir\'e systems, the electron localizations governed by superlattice potentials have attracted particular interest due to their potential to realize strongly correlated physics as well as various quantum device applications~\cite{Andrei2021NatRevMat,Kennes2021NP,Bistritzer2011PNAS,Cao_correlated_2018Nature}. 
In twisted bilayer systems, localized states arise at the centers or vertices of superlattices from moir\'e potentials, 
forming ordered arrays of QDs~\cite{Trambly2010Nanolett,Lopes2012PRB,Wang2018NatNano,Naik2020PRB,zhao2020PRL,Enaldiev2022npj2D,Li2024PRB,Xian2019NL,nakatsuji2025arxiv}. 
A major limitation of bilayer systems is the inherent rigidity of the arrays, as their positions and symmetries are determined by the twist angle. 
Consequently, these configurations remain unaffected by noninvasive external perturbations, such as electric fields, rendering them analogous to conventional QD devices. 
While recent progress in {\it in-situ} mechanical rotation techniques~\cite{Rebeca2018Science,Tang2024Nature} enables partial control over the moir\'e supercell size, this approach could only facilitate the collective displacement of the entire dots in the array. 
So, achieving flexible and reconfigurable local QD patterns as demonstrated in the array of Rydberg atoms~\cite{Bluvstein2022Nature} remains a challenge, prompting the exploration of alternative approaches to unlock new functionalities.

Advances in twisted trilayer systems could provide a chance to overcome this limitation. Recent studies on their moir\'e-of-moir\'e or super moir\'e lattices demonstrate various domain structures with energetically disparate stacking orders, sharply distinguished from the bilayer ones having the energetically equivalent stackings for all domains~\cite{Nakatsuji2023PRX,Craig2024NatMat,VanWinkle2024NatNano,Park2025Nature}. 
Notably, twisted trilayer graphene (TTG) with very small twist angles are shown to realize distinct domain lattices, for which the driving force is the total energy difference between different stackings~\cite{Park2025Nature}. So, if one can control the energy for each domain, their local configurations will alter, presenting a new opportunity to control their electronic properties~\cite{Park2025Nature}.

In this work, we theoretically demonstrate that twisted trilayer hexagonal boron nitride (TTBN) forms well-ordered QD arrays with diverse arrangements and novel reconfigurability based on its ferroelectricity.
For twist angles near 0.1 degrees, TTBNs are shown to generate intricate super moir\'e domain lattices, including triangular, kagome, and hexagram patterns.
The localized quantum states at vertices of these lattices exhibit deep confinement, with binding energies of a few hundred millielectron volts, and display discrete energy eigenstates analogous to those of quantum harmonic oscillators (QHOs). 
Consequently, the uniform QD states could be possible across the whole array provided the superstructure maintains proper orderings~\cite{Nakatsuji2023PRX,Craig2024NatMat,VanWinkle2024NatNano,Park2025Nature}. 

Furthermore, multilayer hexagonal boron nitride (h-BN) can exhibit out-of-plane ferroelectricity, arising purely from broken inversion symmetry induced by the stacking configuration, rather than conventional ionic displacements~\cite{Vizner2021Science, Yasuda2021Science}. 
In bilayer hexagonal boron nitride (BBN), this polarization is directly coupled with external electric fields and results in an ultrafast and reversible change of polarizations through sliding motions of each layer, offering a new platform for robust and switchable electronic functionality~\cite{Yasuda2024,Bian2024}. 
In TTBN, the response is no longer trivial and, depending on the type of tessellation, dynamic repositioning of the QD states at domain vertices is possible.  
This enables seamless local transitions between isolated, capacitively interacting, and tunnel-coupled QDs~\cite{Hsieh2012RPP}, presenting a promising route toward various forms of quantum hybridization.
Our findings suggest that twisted van der Waals (vdW) materials systems hold promise for quantum device applications, including information processing.

\section{Materials and Methods}\label{sec:methods}

\begin{figure}[t]
\centering
\includegraphics[width=1.0\columnwidth]{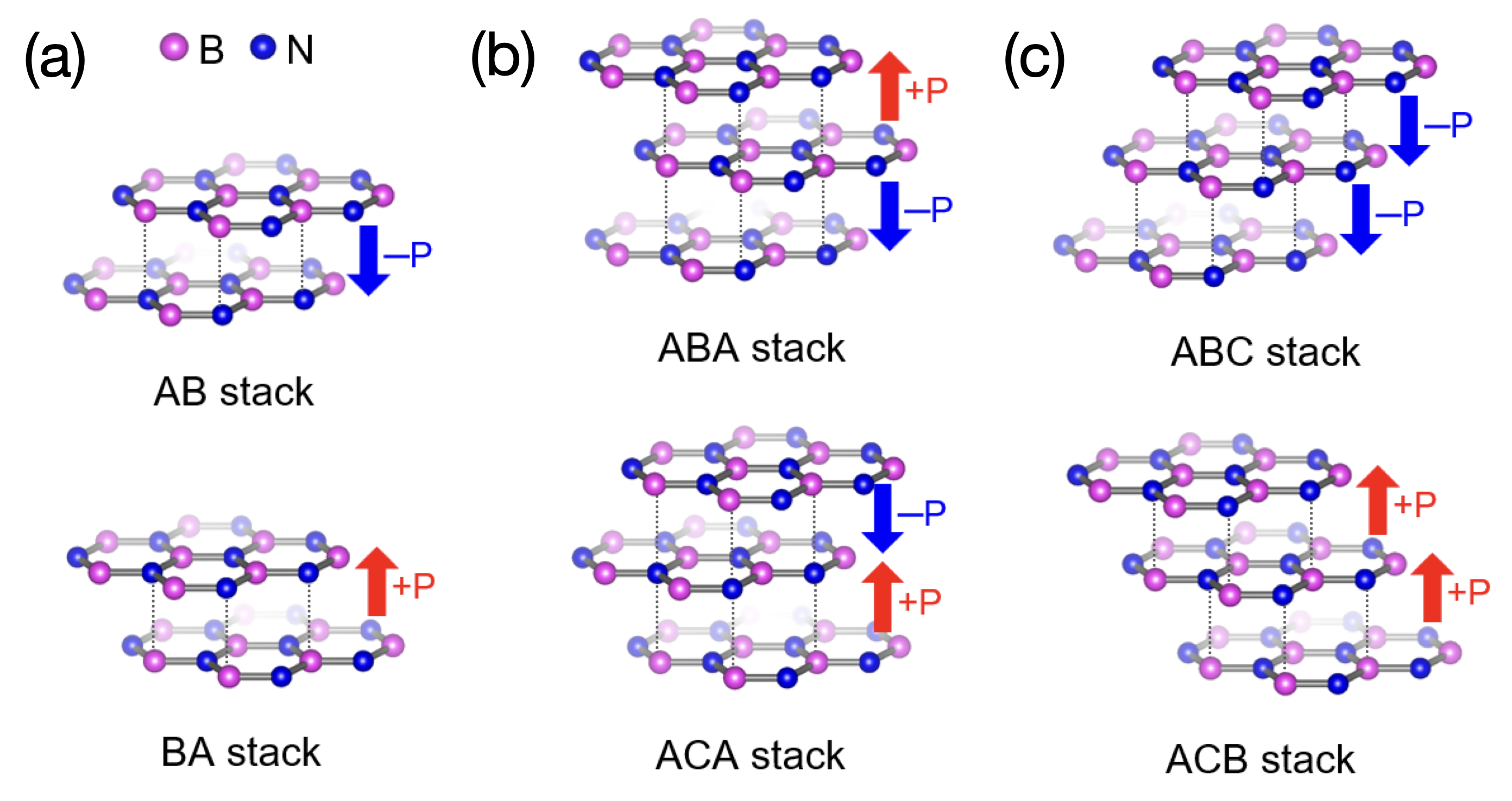}
\caption{Stacking orders of h-BN bilayers and trilayers.
(a) Ball and stick models of bilayer h-BN with AB and BA stackings and their polarizations. 
Trilayer h-BN structures and their polarizations. 
(b) The ABA and ACA stacking orders are nonpolar and (c) ABC and ACB stackings exhibit opposite polarizations. 
}
\label{fig_stacking}
\end{figure}

\subsection{Stacking orders and polarizations}

The h-BN has a honeycomb structure with boron (B) and nitrogen (N) atoms occupying each sublattice.  
Here, for multilayer h-BN, we focus on parallel stack configurations only, where the second layer is positioned directly above the first layer without rotation.
Among the various parallel stackings accessible through in-plane sliding, the AB and BA stackings are most favorable and energetically equivalent~\cite{Constantinescu2013PRL,Warner2010ACSNano}.
In the AB (BA) stacking, boron (nitrogen) atoms in the top layer are positioned directly above nitrogen (boron) atoms in the bottom layer, with the remaining atoms in each layer aligned with vacant hexagon centers as shown in Fig.~\ref{fig_stacking} (a). We note that BA stacking can be assigned as AC stacking also. 
The AB and BA stackings possess opposite out-of-plane polarizations as denoted in Fig.~\ref{fig_stacking} (a)~\cite{Constantinescu2013PRL,Warner2010ACSNano}. 
For the twisted BBN (TBBN), a moir\'e pattern emerges where triangular regions of AB and BA stackings with opposite polarization are arranged in an alternating configuration~\cite{Vizner2021Science, Yasuda2021Science}. 
 
In contrast to TBBN, its trilayer counterpart exhibits four distinct stackings. 
Among these, the nonpolar ABA (ACA) stacking configurations have AB (AC) stacking for the pair of the first and second layers, and BA (CA) stacking for the pair of the second and third layers, respectively. 
As a result, the polarization between the first (bottom) and second (middle) layers cancels the other between the second and third (top) layers, as illustrated in Fig.~\ref{fig_stacking} (b).
The remaining two polar ABC (ACB) stacking are defined by two successive AB (AC) and BC (CB) stackings, thus almost doubling the total polarization. 
The resulting polarizations in ABC and ACB are opposite to each other, as shown in Fig.~\ref{fig_stacking} (c). 

\subsection{Computational methods for lattice relaxations and electronic structures}

We developed and employed a newly designed interatomic potential method~\cite{Park2023NatComm,Park2025Nature} to obtain fully relaxed TTBN structures containing millions of atoms in a commensurate super cell. Note that an incommensurate case can be represented by a nearby commensurate condition without loss of generality~\cite{Park2025Nature}.
Our method achieved excellent agreement with {\it ab initio} calculations while remaining computationally efficient for large-scale systems~\cite{Park2023NatComm}. 
This method integrates harmonic intralayer and long-ranged interlayer potentials, under the assumption that the intralayer potential remains unaffected by the interlayer atomic registry. 
Due to the large in-plane elastic constants of h-BN, the in-plane strain is found to remain below 0.5 \%, justifying the neglect of anharmonic effects. 
The intralayer potential was constructed by fitting to phonon dispersions of monolayer h-BN obtained from density functional perturbation theory (DFPT)~\cite{Baroni2001}, showing excellent agreement with the reference data (see Appendix~\ref{sec:appendix_atom} for detailed information).

To accurately describe interlayer interactions, we extended the Kolmogorov-Crespi (KC) potential by incorporating long-range interactions across all three layers~\cite{Kolmogorov2005PRB,Park2025Nature}. 
While the KC potential provides an accurate description of bilayer interactions, it fails to distinguish between Bernal and rhombohedral h-BN trilayer stackings. 
This interaction was also found to be crucial for small-angle TTG~\cite{Park2025Nature}. 
To resolve this shortcoming, an additional Gaussian-type second-nearest-neighbor interaction between the first and third h-BN layers is introduced, 
enabling our potential model to reproduce the energy differences between ABA and ABC stackings in close agreement with {\it ab initio} benchmarks (see Appendix~\ref{sec:appendix_atom} for detailed parameters of the modified KC potential and the additional long-range potential).

Furthermore, we incorporate the out-of-plane electric polarization of $P_z$
arising between two h-BN layers, which governs the material response to external electric fields. 
To quantitatively capture disparate $P_z$ under various conditions, 
we introduced a pairwise dipole model having both screened and Gaussian-overlap functional forms, 
and parameterized it against {\it ab initio} calculations. 
This model accurately reproduces the dependence of $P_z$
on various stacking arrangements and interlayer spacings, including intermediate configurations between AA and AB stackings, and successfully captures the polarization patterns at polar domain boundaries~\cite{Li2024PRB} of TBBN under applied electric fields 
(see details in Appendix~\ref{sec:appendix_atom}).

To compute the electronic structure, 
we developed an efficient and accurate method based on a tight-binding (TB) approximation. 
The reference electronic structures of few-layer h-BN were calculated using self-consistent extended Hubbard-corrected density functional theory (DFT+$U$+$V$)~\cite{Lee_UV_2020PRR,Rubio2020PRB,Timrov2020prb}. 
This method reproduces the band gap of monolayer h-BN with high accuracy (5.95 eV), 
comparable to the $GW$ approximation~\cite{Liu2022MQT}, and provides on-site ($U$) and intersite ($V$)
Hubbard parameters, which were self-consistently determined for monolayer h-BN and verified to be applicable to bilayer and trilayer systems.
Thus, our calculations are free of band gap errors in typical DFT calculations~\cite{Okada2011PRB}.
Using these results, we obtained Wannier functions constructed from low-energy states using the Wannier90 code~\cite{Pizzi_wannier90_2020}. Details for {\it ab initio} calculations are in Appendix~\ref{sec:appendix_electron}.

The TB model describes the trilayer h-BN using $p_z$
orbitals at each atom as basis states, with intralayer hopping parameters up to seventh-nearest neighbors 
fitted to the DFT+$U$+$V$ Wannier Hamiltonian. 
Interlayer hopping up to next-nearest-layer is parametrized using distance and direction dependent functions, 
reproducing band splittings and detailed band-edge features shown in {\it ab initio} calculations for trilayers.
Additionally, our TB model incorporates on-site energy corrections reflecting local dipole moments arising at the two interfaces depending on the local stacking profile. Effects of external electric fields were simulated by adding a position-dependent potential with a fitted dielectric constant against corresponding first-principles calculations. Detailed parameters are described in Appendix~\ref{sec:appendix_electron}.

\begin{figure}[t]
\centering
\includegraphics[width=1.00\columnwidth]{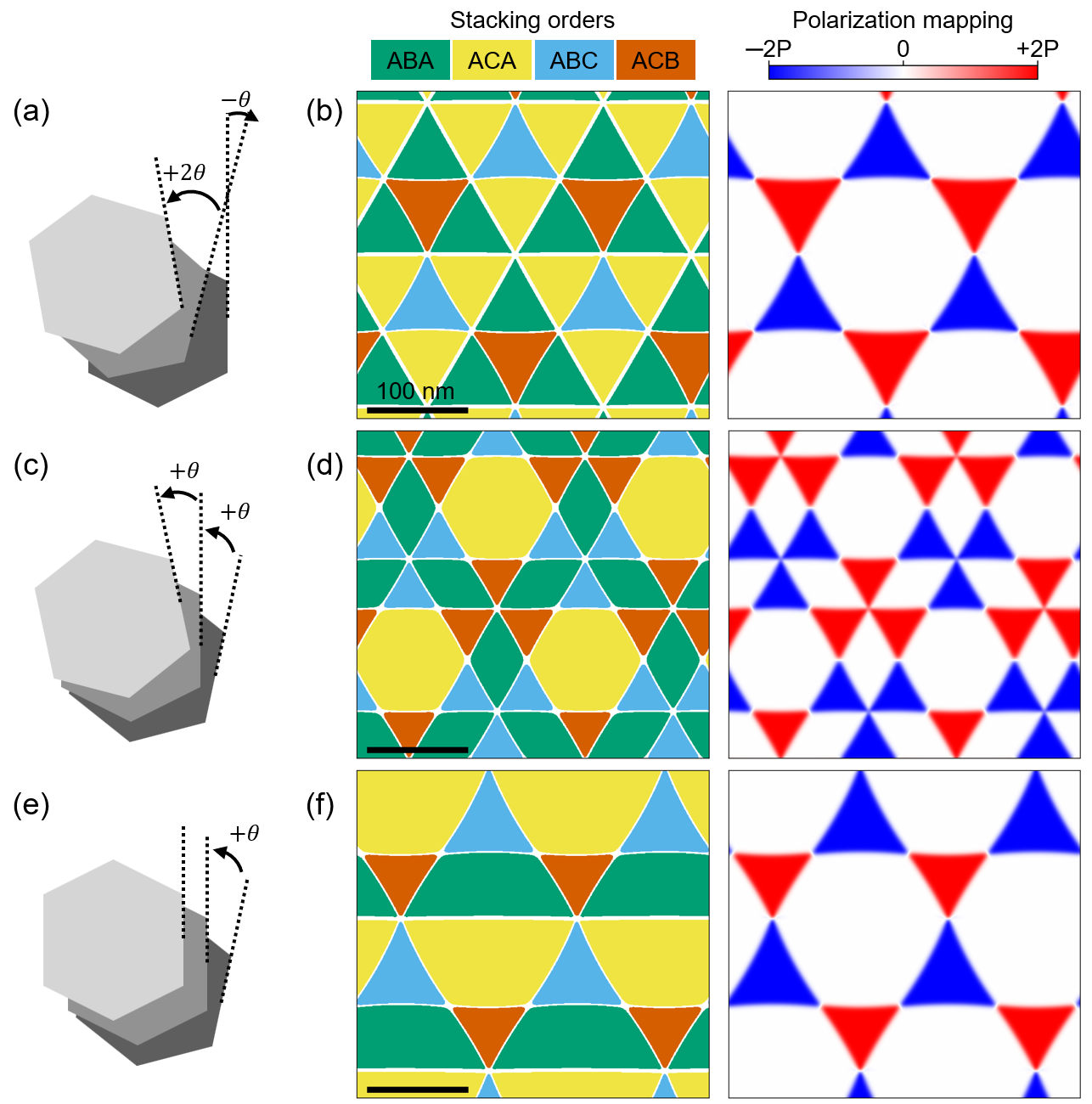}
\caption{
{Super moir\'e domain lattices and associated polarization maps.} 
(a) Twist angles for alternate stacking TTBN with $\theta_{12}=-\theta$ and $\theta_{23}=2\theta$, and 
(b) its relaxed structure for a case of $\theta=0.082^\circ$ showing various stacking domains (four different colors) in the left panel and its associated polarization distribution in the right panel.
Red and blue colored polarizations follow directions in Fig.~\ref{fig_stacking} (c). The black scale bar is 100 nm. 
(c) and (d), Twist angles, relaxed structures and polarization distributions for helical stacking with $\theta_{12}=\theta_{23}=\theta=0.082^\circ$. 
(e) and (f), Similar displays for a twisted monolayer-bilayer with $\theta=0.082^\circ$.
}
\label{fig_domain}
\end{figure}

\section{Results}\label{sec:results}

\subsection{Unconventional super moir\'e lattices and polarization domains}

Among the various trilayer stacking orders without twist shown in Fig.~\ref{fig_stacking}, 
the nonpolar stacking configurations (ABA and ACA) exhibit a slightly lower total energy, less than 0.5 meV per atom, compared to the polar stackings (ABC and ACB)~\cite{Haga2021}.
Due to the competing stacking orders, TTBN with small twist angles can exhibit a variety of super moir\'e lattices, as illustrated in Fig.~\ref{fig_domain}. 
Similar to prior studies on TTG~\cite{Park2025Nature}, as the twist angles decrease to near 0.1$^\circ$, the energy gain from increasing the area of the lowest-energy stacking configuration can surpass the energy cost associated with the growing boundaries between domains. 
In contrast to marginally TTG, however, TTBN structures not only feature domain lattices with varying stackings but also possess an associated network of polarization domains, which may not precisely follow the pattern of the stacking domains as will be discussed below.

Considering various domain shapes, 
we will focus on three representative ones having characteristic sets of twist angles.
The first case is the alternate stacking where the twist angle ($\theta_{12}$) between the first and second layer has an opposite direction with respect to the twist angle ($\theta_{23}$) between second and third layer as shown in Fig.~\ref{fig_domain} (a). For a specific case with $\theta_{23}=-2\theta_{12}$, the relaxed super moir\'e lattice shows a tessellation of triangular domains with different stackings, as shown in Fig.~\ref{fig_domain} (b), which consists of hexagonal arrangements of alternating triangular domains of ABA and ACA stackings surrounded by triangular domains of ABC and ACB stackings. Because of the nonpolar nature of ABA and ACA orders, the associated polarization domain shows a kagome-shaped distribution of alternating out-of-plane directions as displayed in Fig.~\ref{fig_domain} (b). 

Second, we will consider an opposite twist geometry to the first one, where two twist angles have the same direction. 
In the case of helical stacking with $\theta_{12}=\theta_{23}$ shown in Fig.~\ref{fig_domain} (c), 
the relaxed structure shows a network of corner-shared hexagrams as shown in Fig.~\ref{fig_domain} (d), being similar to the TTG case~\cite{Park2025Nature}.  
The hexagonal area of ACA stacking and rhombic area of ABA stacking are nonpolar so that the resulting polarization distribution displays an alternating set of triangular domain clusters with opposite polarization directions as shown in Fig.~\ref{fig_domain} (d) and that all domain clusters with opposite polarizations are connected through their vertices.

\begin{figure}
\centering
\includegraphics[width=1.0\columnwidth]{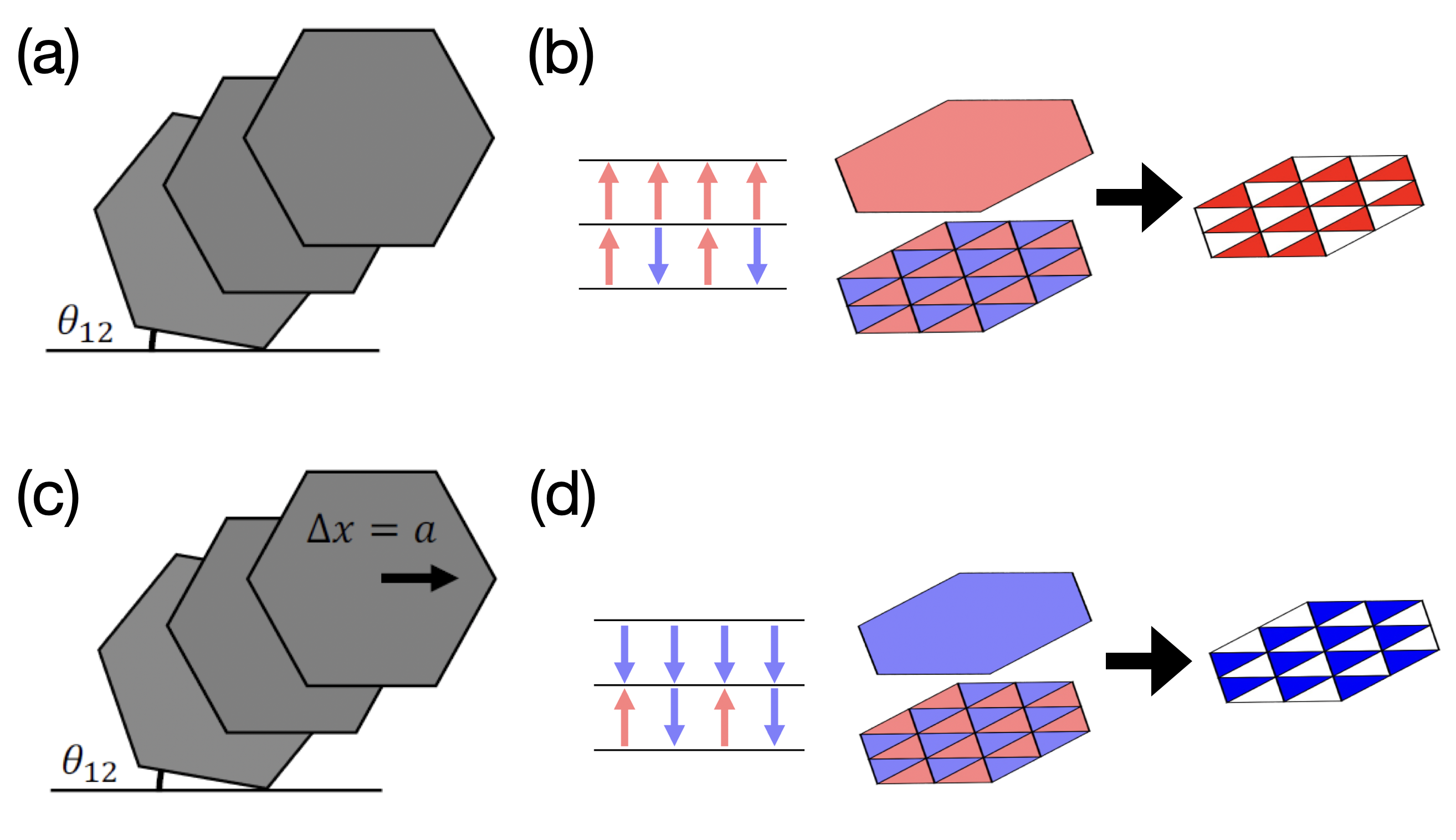}
\caption{
{Polarization domains of twisted monolayer-bilayer system with moderate twist angles ($\theta_{12}>0.1^\circ$ and $\theta_{23}=0$).} 
(a) BA-stacked and (c) AB-stacked bilayer on top of monolayer with twist angle $\theta_{12}$. The twist geometry in (c) can be obtained from a uniform sliding of the top layer in (a) by $\Delta x =a$ ($a$ is the B–N bond length).
The corresponding polarization domains are drawn in (b) and (d), respectively. 
The red, blue and white correspond to upward, downward and zero polarizations, respectively.
In the left, middle, and right panels, cross sectional, pair-wise layer-resolved and total polarizations are displayed. 
See the text for a detailed description. 
}
\label{efig_TwistMonoBi}
\end{figure}

The third configuration is the twisted monolayer-bilayer, a bilayer ($\theta_{23}=0$) on top of a monolayer with a twist by $\theta_{12}>0$ as shown in Fig.~\ref{fig_domain} (e).
This twist geometry displays two distinguishable domain tessellations. 
Figure~\ref{fig_domain} (f) shows one of the two domains, the kagome-like twisted monolayer-bilayer, which has undergone a spontaneous symmetry breaking at a smaller twist angle~\cite{Park2025Nature}. 
Here, being similar to the case in TTG~\cite{Park2025Nature}, the hexagonal Bernal stacked region is divided into ABA and ACA stackings by a nematic domain boundary, which is surrounded by alternating rhombohedral stacked ABC and ACB triangular domains, thus forming the kagome-like domain tessellation.  
As will be discussed later, this is a metastable structure that could be stabilized at very small twist angle. 
In a moderately small twist angle ($\theta>0.1^\circ$), the twisted monolayer-bilayer prefers the simpler triangular domain network composed of alternating polar and nonpolar regions, as illustrated in Fig.~\ref{efig_TwistMonoBi}, which yields a net polarization in the system.
The related unique sliding ferroelectricity in this geometry will be discussed below. 

\begin{figure*}
\centering
\includegraphics[width=0.95\textwidth]{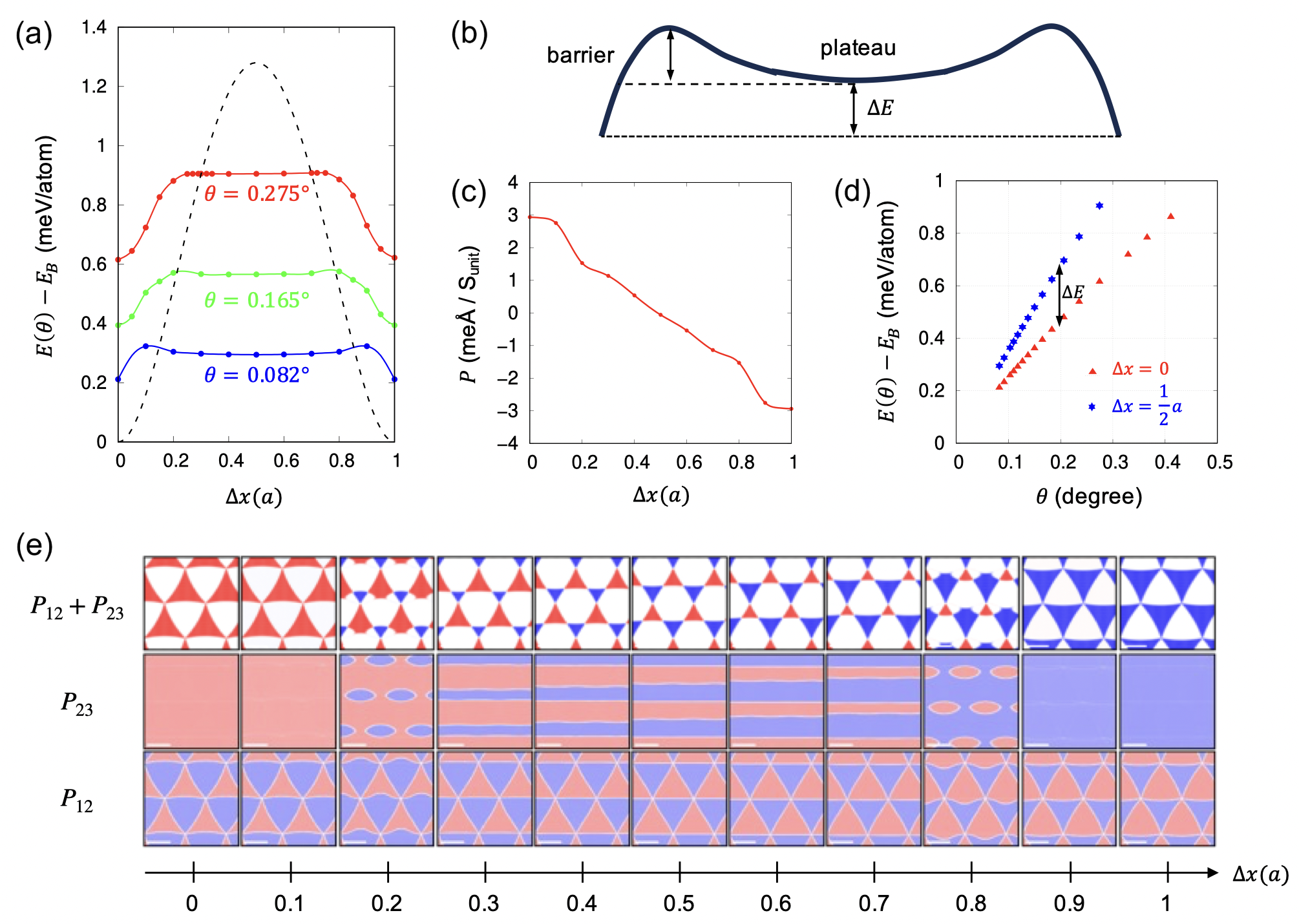}
\caption{{Sliding ferroelectricity and energy landscape of twisted monolayer-bilayer hBN.}
(a) 
Energy landscapes as a function of the translation $\Delta x$ (in units of B–N bond length $a$) of the top layer in twisted monolayer–bilayer h-BN. 
Three twist angles are considered. The dashed line represents the sliding energy landscape of the bilayer BN system. 
For comparison, the energies for each twist angle are vertically shifted such that the zero energy of each colored curve corresponds to its respective intercept with the energy axis.
(b) Schematic energy landscape from (a) showing a plateau around  $\Delta x =a/2$.
(c) Polarization switching during the translation and zero net polarization occurs at $\Delta x = a/2$. 
(d) The energy difference $\Delta E$ between the triangular type ($\Delta x = 0 $) and kagome-like domains ($\Delta x =a/2$). 
(e) 
Layer-resolved interlayer polarization densities as a function of top-layer sliding by $\Delta x$.
$P_{12}$ ($P_{23}$) indicates the interlayer polarization density between the bottom and middle (middle and top) layers.  Top panels are variation of total polarizations as a function of translation ($\Delta x$) of the top layer.
}
\label{efig_Sliding}
\end{figure*}

\subsection{Sliding ferroelectrics in twisted monolayer-bilayer moir\'e systems}

When a Bernal-stacked BBN is placed on top of a monolayer with a moderate twist angle, a triangular moir\'e domain forms, similar to twisted bilayer systems.
In these types of stackings, the net polarization density is the sum of two interlayer polarizations: one from the twisted monolayer–bilayer interface and the other from the Bernal-stacked bilayer itself, as shown in Fig.~\ref{efig_TwistMonoBi}.

Since a lateral translation of one layer by $\Delta x= a$ ($a$ is a nearest-neighbor  bond length between B and N atoms) can reverse the polarization direction as shown in Fig.~\ref{fig_stacking} (a), even the same moir\'e patterns may display the opposite polar direction.
In the left panels of (b) and (d) in Fig.~\ref{efig_TwistMonoBi}, the red and blue arrows denote schematic polarization between the adjacent layers. For BA (AB)-stacked BBN for the top and middle layers, it has a uniform polarization domain (also denoted as uniform red and blue hexagons in the middle panels) while the alternating red and blue triangular domains denote polarization map between the middle and bottom layers with a twist angle of $\theta_{12}$. In the right panels, by adding up two polarization domains, the final patterns show an interesting switching behavior:  
the sliding top layer by $a$ causes the previously polarized domains to be unpolarized and vice versa, as shown in Figs.~\ref{efig_TwistMonoBi} (c) and (d).

\begin{figure*}
\centering
\includegraphics[width=0.75\textwidth]{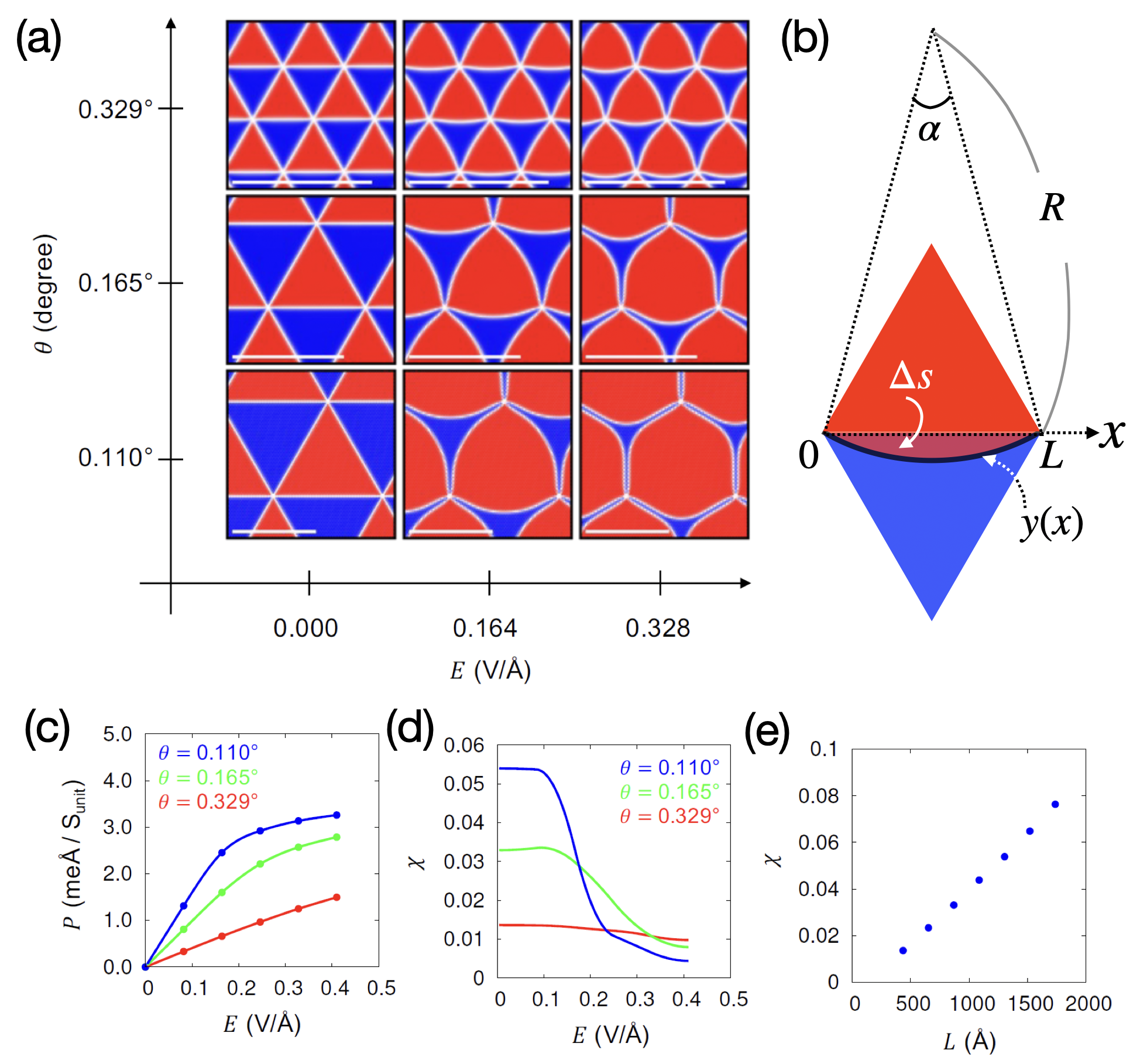}
\caption{
(a) Out-of-plane polarization densities of TBBNs with twist angles of $\theta=0.110^\circ, 0.165^\circ~\text{and}~0.329^\circ$ with increasing electric fields ($E$). The scale bar (white line) is 100 nm.
The field direction is (anti)parallel to the polarization direction of BA (AB) stacking (red and blue).
(b) Schematic geometry of domain wall under the external electric field. $L$ represents the length of the triangular domain wall and $\Delta s$ is the increased area of the triangular domain by a warped boundary under $E$. The warped boundary (thick black line) is described by $y(x)$ along the $x$-axis denoted by a dotted black line. The warped boundary has an arc shape that can be defined as a segment of a circle with radius $R$, subtended by a central angle $\alpha$.
(c) Net polarization densities for the three cases as a function of $E$. 
(d) Electric susceptibilities $\chi$ of the three cases as a function of $E$. 
(e) The computed electric susceptibility of $\chi$ as a function of $L$ for the fully relaxed TBBNs under electric fields using our interatomic potential method.
}
\label{efig_BBN}
\end{figure*}

Due to the polarization reversal by the translation, we expect that an external electric field could potentially switch between these two configurations. However, unlike a previous demonstration of sliding ferroelectric behavior in bilayer studies~\cite{Yasuda2021Science,Vizner2021Science}, the TTBN has competing Bernal and rhombohedral stacking orders so that the locally stable moir\'e domain other than a simple triangular one may be possible.

To quantify the switching mechanism, we calculated the energy barrier between the two polarization states as the top layer slides.
Figure~\ref{efig_Sliding} (a) shows the barrier profiles for the twisted monolayer-bilayer with $\theta_{12}=0.082^\circ$, 0.165$^\circ$ and 0.275$^\circ$, compared with bilayer sliding  (dashed line).
Notably, unlike the maximum energy at $\Delta x= a/2$ for sliding BBN, the full atomic relaxation significantly lowers the energy barrier in trilayer cases compared to rigid sliding, resulting in a broad plateau with a shallow minimum at $\Delta x = \frac{1}{2}a$, as is clearly shown for the case with $\theta_{12}=0.082^\circ$.
The corresponding energy landscape is schematically illustrated in Fig.~\ref{efig_Sliding} (b). 

Figure ~\ref{efig_Sliding} (e) displays the layer-by-layer polarization densities along the sliding path, revealing that most of the change arises from $P_{23}$ (polarization between the middle and top layers), while $P_{12}$ remains relatively constant.
When atoms are fixed during sliding, the energy profile should resemble that of bilayer sliding. 
With relaxation, however, the system can lower its energy by expanding low-energy stacking domains at the cost of in-plane strain.
Because stacking energy scales with area ($\propto L^2$), while strain energy scales with length ($\propto L$), at small twist angles, all areas tend to be occupied by pure AB and AC domains. 
The area ratios of the AB and AC domains in $P_{23}$ are approximately $\frac{\Delta x}{a}$ and $1-\frac{\Delta x}{a}$, respectively.
At $\Delta x = \frac{1}{2}a$, $P_{23}$ is partitioned into equally spaced AB and AC stripes as shown in Fig.~\ref{efig_Sliding} (e).
Deviations from this point introduce energy costs due to weak interwall interactions and result in a very shallow minimum around $\Delta x = \frac{1}{2}a$. 
If the two walls are too close ($\Delta x = 0.2a$ or $\Delta x = 0.8a$), the stripes break into small ellipses to reduce the interwall energy, manifesting as small humps in the energy landscape in Fig.~\ref{efig_Sliding} (b).

The net polarization density along the sliding path is shown in Fig.~\ref{efig_Sliding} (c). 
At $\Delta x =\frac{1}{2}a$, a kagome-like domain pattern emerges as a local minimum energy configuration and maximizes nonpolar domains.
As shown in Fig.~\ref{efig_Sliding} (d), the energy difference $\Delta E$ between the triangular type and the kagome-like domains decreases with decreasing twist angle, due to the increasing importance of the stacking energy relative to the energy of the domain wall.
This suggests that at sufficiently small twist angles ($\ll 0.1^\circ$), the kagome domain structure could be the true ground state of the twisted monolayer-bilayer BN.

\subsection{Electric field control of super moir\'e domains}

\begin{figure*}[t]
\centering
\includegraphics[width=1.0\textwidth]{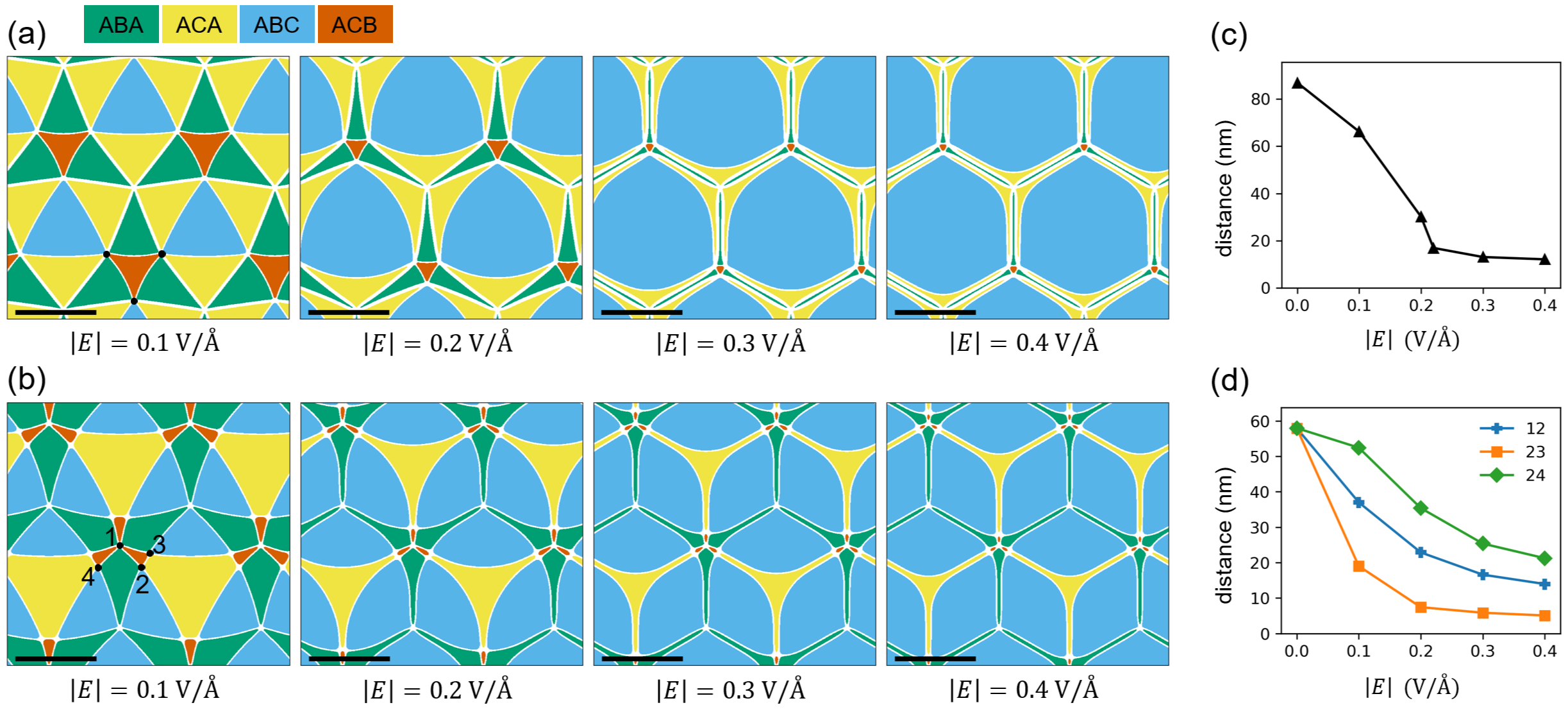}
\caption{
The field-induced ($E < 0$) evolution of domain structures of TTBNs with (a) alternate and (b) helical stacking configurations, corresponding to Figs.~\ref{fig_domain} (b) and (d), respectively. 
The scale bar is 100 nm.
(c) and (d) The changes in the distances between boundary vertices that host quantum dot states. The dot positions are indicated in the leftmost panels of (a) and (b) by black dots. In (c), the distance measures the side of equilateral triangles formed by three dots, whereas in (d) the distances vary, as labeled by the two dot indices shown in (b).
}
\label{fig_domain_efield}
\end{figure*}

At small twist angles, the local domain lattice within each moir\'e supercell forms to maximize the lowest-energy stacking configuration. As a result, domain tessellation strongly depends on the relative energies of competing stackings and can vary if these energies change~\cite{Park2025Nature}. Marginally twisted BBNs follow the same principle, with the added feature of polarization domains. 
In TBBN~\cite{Li2024PRB,Xian2019NL}, the most favorable stackings of AB and BA (or equivalently AC) are energetically degenerate. But if we apply the out-of-plane electric field $E$, the domains with the parallel polarization to $E$ become energetically favorable, and the domain walls become warping to increase the area of the domains at the expense of additional energy cost from the increased length of the domain walls. Figure~\ref{efig_BBN} (a) shows the polarization density map of TBBN and their evolutions in the external electric fields for three different twist angles. We note that the side length of the triangular moir\'e domain is inversely proportional to the twist angle $\theta$. We also notice that the vertices of the triangular domain are all fixed for a given twist angle irrespective of the strength of $E$.

Figure~\ref{efig_BBN} (c) shows the net polarization density of $P$ per the moir\'e cell with increasing $E$. 
As $\theta$ decreases, $P$ increases for a given $E$. 
The electric susceptibilities $\chi=\varepsilon_{r}-1$ ($\varepsilon_{r}$: relative permittivity) are plotted in Fig.~\ref{efig_BBN} (d). 
If we neglect the interaction between the walls, $\chi$ has a simple relation with the energy of the domain wall. 
Let us assume that $E$ makes the energy difference $\delta_E$ between the AB and AC domains.  
For $\delta_E \neq 0$, the domain wall becomes warped, and the energy $F$ of the system using the geometry shown in Fig.~\ref{efig_BBN}(b)
can be written as
\begin{equation}
F = \int_0^L dx\, E_W \sqrt{\left( y'(x) \right)^2 + 1} + \int_0^L dx\, \delta_E\, y(x),
\end{equation}
where $E_W$ is the energy of the domain wall per unit length. 
By applying the Euler-Lagrange equation, we obtain the following.
\begin{equation}
\frac{d}{dx} \left( \frac{y'}{\sqrt{(y')^2 + 1}} \right) = \frac{\delta_E}{E_W}.
\label{Eq:euler}
\end{equation}
Because the left-hand side in Eq.~\eqref{Eq:euler} is the curvature of the domain wall, the shape of the domain wall is the arc of a circle with a radius $R = \frac{E_W}{\delta_E}$.
The increased area of $\Delta_S$ can be approximated for small $E$ as 
$
\Delta_S = \frac{R^2 \alpha}{2} - \frac{R L}{2} \cos\left( \frac{\alpha}{2} \right) \simeq \frac{L^3}{16R}
$ in the case of $R \gg L $. 
Then, the increased $P$ upon warping can be written as 
$
P = \frac{4 \sqrt{3}p\Delta_S}{L^2} \simeq \frac{\sqrt{3}}{4}  \frac{\delta_E}{E_W} L p = \frac{\sqrt{3}}{2}  \frac{L p^2}{E_W} E
$
where $p$ is the polarization density per unit area and $\delta_E=2pE$. 
Then, the electric susceptibility can be written as 
\begin{equation}
\chi = \frac{P}{\varepsilon_0 E} = C_\chi  \frac{L}{E_W}, 
\label{Eq:sus}
\end{equation}
where $C_\chi=0.0024672$ and the units of $L$ and $E_W$ are \AA{} and meV/\AA{}, respectively.   
We confirm that the linear relationship between $\chi$ and $L$ in Eq.~\eqref{Eq:sus} is numerically verified for $E=0.08$~V/\AA~using our interatomic potential method as shown in Fig.~\ref{efig_BBN} (e) and the estimated energy of the domain wall is 56.5 meV/\AA. 

Under external electric fields, the moir\'e domains in TBBN have been shown to deform, with their boundaries becoming distorted, consistent with previous experimental and theoretical studies~\cite{Weston2022NatureNano,Li2024PRB}. However, it is important to note that both the supercell lattice and the positions of the domain vertices are determined solely by the twist angle. Reversing the direction of the applied electric field does not alter these structural features; it merely switches the polarization~\cite{Vizner2021Science,Yasuda2021Science}. Consequently, precise spatial control over the domain symmetries and the associated local quantum states at the domain vertices of twisted bilayer systems are hardly realizable.

Unlike the twisted bilayer cases discussed so far, the trilayer systems show markedly different behaviors under external electric fields. 
In TTBN under a homogeneous perpendicular electric field, 
the local domain walls distort to increase the domain area where its polarization is parallel to the field, as was demonstrated in TBBN under the field. 
In contrast to TBBNs, all the other domains surrounding the expanded domains start to contract with the field and all the other boundaries become shortened as shown in Fig.~\ref{fig_domain_efield}. 
In the case of alternate stacking displayed in Fig.~\ref{fig_domain} (b), the area of equilateral triangular domains corresponding to ACB stackings decreases significantly with increasing field and eventually enters a saturation regime at 0.22 V/\AA, as depicted in Figs.~\ref{fig_domain_efield} (a) and (c). 
Its fully saturated side length is approximately 10 nm at $E>0.3$\ V/\AA. This saturation in domain size arises from interactions between domain boundaries and vertices, which limit further shrinkage. We note that this highly nonlinear variation of domain area with electric field follows a 
single, well-defined scaling rule, unifying all calculated results across varying twist angles, as discussed in
Appendix~\ref{sec:appendix_QD_scale}. We also find that the critical field required to reach this saturation remains well below the dielectric breakdown threshold of trilayer h-BN, larger than 0.8 V/\AA~\cite{Okada2011PRB}.

In the case of helical stacking, illustrated in Fig.~\ref{fig_domain} (d), a similar trend is observed: The isosceles triangular ACB stacking domains contract rapidly with increasing  field and eventually saturate at a shortest side length of approximately 5 nm, as shown in Figs.~\ref{fig_domain_efield} (b) and (d).
The polarization distributions shown in Fig.~\ref{efig_Pol_Efield} in Appendix~\ref{appendix:slideFE} confirm this trend, demonstrating that the regions with polarization aligned in the direction of the applied field grow in area, consistent with energetic favorability under the external field.

The metastable kagome-shaped domain pattern of the twisted monolayer-bilayer TTBN exhibits minimal domain wall distortion under applied external electric fields as shown in Fig.~\ref{efig_Field_MonoBi}. Energetically favorable domains expand primarily through domain wall motion, and the system undergoes a structural transition into a simple triangular domain configuration even under a weak external field of 0.04 V/\AA, as elaborated in Appendix~\ref{appendix:slideFE}. Moreover, reversing the direction of the electric field alters only the polarization direction within each triangular domain, without shifting the QD-hosting vertex position itself. Thus, for the twisted monolayer-bilayer case, the electric-field-induced switching introduces a new sliding ferroelectric transition between kagoma-shaped to triangular-shaped polarization domain lattices. 

We note that the striking difference between TBBN and TTBN arises from the presence of competing stacking domains within the supermoir\'e periodicity of trilayer systems. In TTBN, multiple domains with distinct stacking orders coexist within a single superlattice unit cell defined by two twist angles. As a result, the domain boundaries are not rigidly fixed within the supercell, allowing external electric fields to modify the local arrangement of the supermoir\'e domain tessellation and the positions of the associated vertices.

\begin{figure*}[t]
\centering
\includegraphics[width=1.0\textwidth]{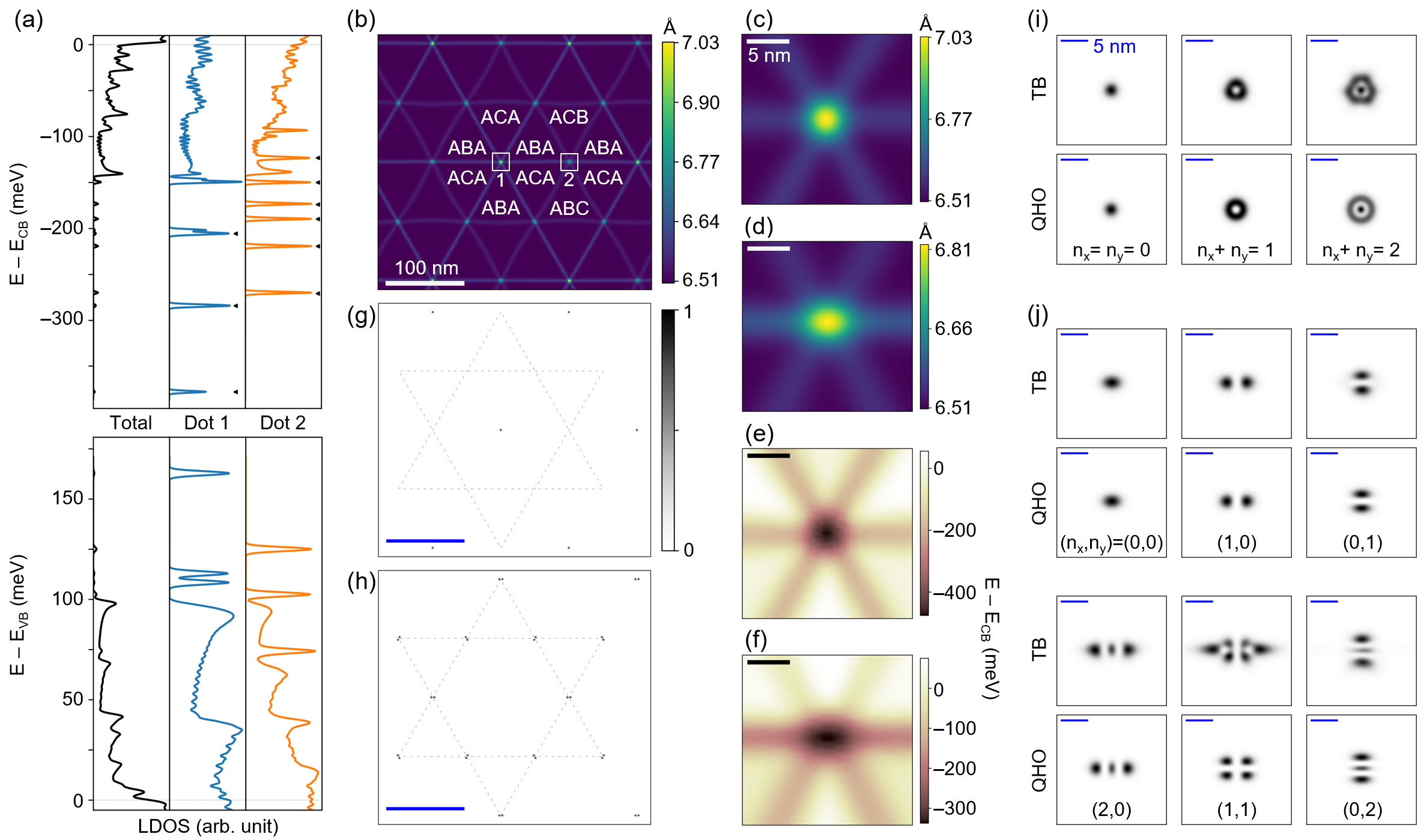}
\caption{
{Localized quantum harmonic oscillator states in TTBN.} 
(a) Electronic structure of alternate stacking geometry with $\theta_{12}=-0.082^\circ$ and $\theta_{23}=0.164^\circ$ shown in Fig.~\ref{fig_domain} (b). The total density of states (denoted as ``Total'') and local density of states (LDOS) for rectangular sites 1 and 2 in (b) (denoted as ``Dot 1'' and ``Dot2'') are shown for the conduction bands and valence bands (top and bottom panels). The energy level is a relative value to the conduction and valence band edges ($E_\text{CB}$ and $E_\text{VB}$) of the bulk trilayer stackings at the center of triangular domains. 
(b) The fully relaxed interlayer spacing between the bottom and top layers drawn for the same region as Fig.~\ref{fig_domain} (b). 
(c) and (d) The enlarged interlayer spacings of Dot 1 and Dot 2 in (b) are shown, respectively. 
(e) and (f) The corresponding potential wells of conduction band energy. 
(g) and (h) The spatial distributions of the lowest eigenstates $|\psi|^2$ at Dot 1 and the second lowest state at Dot 2, respectively. The colormap range is normalized to the maximum value of each panel. The dashed lines are visual guides for the polar domain boundaries. 
(i)
The enlarged spatial densities of the first three lowest states for Dot 1, of which energies are indicated by small black triangles in (a). 
The upper panels are from our TB calculations, denoted as TB and the lower ones from 2D QHO potentials denoted as QHO with corresponding quantum numbers of $n_x$ and $n_y$. 
(j) Same as (i) for Dot 2 with the six lowest states of which energies are also marked by the triangles in (a). 
}
\label{fig_alt_qho}
\end{figure*}

\subsection{Realization of quantum dot arrays}

As we mentioned earlier, several twisted bilayer systems including TBBN can host arrays of localized states at the center or vertices of their moir\'e lattice potentials~\cite{Trambly2010Nanolett,Lopes2012PRB,Wang2018NatNano,Naik2020PRB,zhao2020PRL,Enaldiev2022npj2D,Li2024PRB,Xian2019NL,nakatsuji2025arxiv}. 
For marginal TTBNs, we also expect similar but distinct localized states. 
In Fig.~\ref{fig_alt_qho} (a),  total density of states (TDOS) of the alternate stacking case in Fig.~\ref{fig_domain} (b) is drawn. 
The local densitiy of states (LDOS) for two distinguishable vertices [1 and 2 in Fig.~\ref{fig_alt_qho} (b), which will be referred to as Dot 1 and Dot 2] are also displayed. 
Below the energy of the conduction band edge ($\text E_\text{CB}$) and above the valence band edge ($\text E_\text{VB}$), several discrete energy levels from localized states are identified. Here we define $\text E_\text{CB}$ and $\text E_\text{VB}$ as the corresponding band extrema obtained at the centers of triangular domains.

We found that the origin of confinements for these discrete states is the spatial variation of band energy around vertices depending on their stacking orders. 
In Fig.~\ref{fig_alt_qho} (b), the spatial variation of interlayer distances between top and bottom layers are drawn. 
Near Dot 1, its enlarged view in Fig.~\ref{fig_alt_qho} (c) shows threefold rotational symmetric variations circling around the isotropic maximum area. 
Near Dot 2, the corresponding distance map shows twofold symmetry around the elongated maximum region as shown in Fig.~\ref{fig_alt_qho} (d). It turns out that each area for interlayer distance maximum corresponds to unstable AAA stacking and mixture of AA and SP stackings for Dots 1 and 2, respectively. The symmetry differences in Figs.~\ref{fig_alt_qho} (c) and (d) arise from the surrounding stackings. Dot 1 is encircled by nonpolar ACA and ABA stackings with three-fold symmetry, while Dot 2 is surrounded by nonpolar ACA and ABA stacking together with polar ACB and ABC stackings that break this symmetry as shown in Fig.~\ref{fig_alt_qho} (b).
Then, we plot the energy of conduction band minima from energy bands for corresponding local stackings in Figs.~\ref{fig_alt_qho} (e) and (f). It is noticeable that both energy maps follow the distance profiles in an opposite way, and show deep isotropic and anisotropic potentials for Dot 1 and Dot 2, respectively.
These smooth potentials prompt us to consider two-dimensional (2D) QHO potential,
$
V_\text{Dot}(x,y)=\frac{1}{2}m_x\omega_x^2 x^2 +\frac{1}{2}m_y \omega_y^2 y^2,
$
where $m_x$ and $m_y$ are effective masses and $\omega_x$ and $\omega_y$ are angular frequencies. The four parameters are fit to generate the low energy landscapes of each dot. The resulting parameters for the QHO potentials for Dot 1 and Dot 2 are summarized in Table~\ref{etable_qho_parameter}. We note that for Dot 1, the fitted parameters satisfy $m_x = m_y$ and $\omega_x = \omega_y$, while for Dot 2, anisotropy is necessary.

In Figs.~\ref{fig_alt_qho} (g) and (h), we present the wavefunctions of the lowest eigenstate localized at Dot 1 and the second-lowest one at Dot 2, respectively, to illustrate their spatial distributions. The eigenstate associated with Dot 1 is localized at the centers of the nonpolar hexagons within the kagome-shaped polar domain lattice, whereas the state associated with Dot 2 is at the vertices of these hexagons.
To further investigate the characteristics of the localized states, we plot the wavefunctions of the three lowest eigenstates for Dot 1 and the six lowest eigenstates for Dot 2 in Figs.~\ref{fig_alt_qho} (i) and (j), respectively. Simultaneously, we compute the corresponding eigenstates of the effective 2D Schr\"odinger equation with QHO potential using parameters listed in Table~\ref{etable_qho_parameter}, and we compare these with the results obtained from full-scale TB simulations involving multimillion atoms. 
The 2D QHO model yields eigenenergies of 
$E_\text{Dot1}=\hbar\omega(n_x+n_y+1)$  for Dot 1 and $E_\text{Dot2}=\hbar\omega_x(n_x+\frac{1}{2})+\hbar\omega_y(n_y+\frac{1}{2})$ for Dot 2 where $n_x,n_y=0,1,2\cdots$. 
The eigenstates from our TB calculations closely match those of the 2D quantum harmonic oscillator (QHO) in both spatial wavefunction profiles and energy level structures, characterized by quantum numbers of $n_x$ and $n_y$. 
Detailed energy comparisons with the 2D QHO model are summarized in Table~\ref{etable_qho_energy}. Minor deviations appear in the second and third eigenstates of Dot 1, and the fourth and fifth of Dot 2, due to boundary-induced perturbations.

\begin{table}[t]
\caption{ {The parameters of 2D QHO potentials for Dot 1 and Dot 2 in Fig.~\ref{fig_alt_qho}.} CB (VB) stands for QHO states below (above) the conduction (valence) band edge. 
The unit for $m_x$ and $m_y$ is the electron mass $m_e$. 
$E_0$ is the zero level of the harmonic potential with respect to the bulk reference level. The unit of $\hbar\omega_x$, $\hbar\omega_y$ and $E_0$ is meV.}\label{etable_qho_parameter}
\centering
\begin{ruledtabular}
\begin{tabular}{lccccc}
  & $m_x$  & $m_y$  & $\hbar\omega_x$  & $\hbar\omega_y$  & $E_0$ \\
\hline
Dot 1 (CB) & 0.451    & 0.451    & 103 & 103 & $-471$\\
Dot 1 (VB) & $-0.616$ & $-0.616$ &  66 &  66 &  222\\
Dot 2 (CB) & 0.443    & 0.523    &  54 &  92 & $-339$\\
Dot 2 (VB) & $-0.590$ & $-0.691$ &  29 &  65 &  169\\
\end{tabular}
\end{ruledtabular}
\end{table}

\begin{table}[b]
\caption{{Energies of QD states with quantum numbers $(n_x,n_y)$ using TB approximation and 2D QHO models.}
Relative values to the conduction band (CB) edge $E-E_{\text{CB}}$ and to the valence band (VB) edge $E-E_{\text{VB}}$ in meV.}\label{etable_qho_energy}
\centering
\begin{ruledtabular}
\begin{tabular}{lcccccccc}
$(n_x,n_y)$  & $(0,0)$ & $(1,0)$ & $(0,1)$  & $(2,0)$ & $(1,1)$  & $(0,2)$ \\
\hline
Dot 1 (CB) TB     & $-378$ & $-284$ & $-284$ & $-204$ & $-204$ & $-204$ \\
Dot 1 (CB) QHO    & $-368$ & $-265$ & $-265$ & $-162$ & $-162$ & $-162$ \\
\hline
Dot 1 (VB) TB     & 163 & 111 & 111 & - & - & - \\
Dot 1 (VB) QHO    & 155 & 89 & 89 & - & - & - \\
\hline
Dot 2 (CB) TB     & $-270$ & $-219$ & $-190$ & $-173$ & $-150$ & $-123$ \\
Dot 2 (CB) QHO    & $-266$ & $-212$ & $-174$ & $-158$ & $-120$ & $-81$ \\
\hline
Dot 2 (VB) TB     & 125 & 102 & - & - & - & - \\
Dot 2 (VB) QHO    & 121 & 93 & - & - & - & - \\
\end{tabular}
\end{ruledtabular}
\end{table}

For the helical stacking configuration [Fig.~\ref{fig_domain} (d)] and the twisted monolayer–bilayer system [Fig.~\ref{fig_domain} (f)], electronic structures (TDOS and LDOS), spatial maps of interlayer distances, corresponding potential landscapes, and 2D QHO type eigenstate wavefunctions are shown in Figs.~\ref{efig_Helical_QHO} and \ref{efig_MonoBi_QHO} in Appendix~\ref{sec:appendix_QD_others}, respectively. 
Associated parameters for these QHO potentials and energies of QD states are also summarized in Tables~\ref{etable_qho_parameter2} and~\ref{etable_qho_energy2} in Appendix~\ref{sec:appendix_QD_others}.
Furthermore, the QD states and their energy level spacings are robust under a moderate inhomogeneity such as twist angle and strain disorders (see Appendix~\ref{sec:appendix_inhomo}).
These results confirm that TTBN with varying stacking geometries can host robust arrays of localized 2D QHO states.

In addition to the regular arrays of deep QD states discussed so far, we note the distinct one-dimensional (1D) features in the DOS between the discrete QD levels and the bulk band edges, as shown in Fig.~\ref{fig_alt_qho} (a).
These 1D-like DOS features are observed in both the alternate and helical stacking configurations and are attributed to boundary states localized at the interfaces between distinct stacking domains. 

To elucidate the nature of these boundary states, the real-space distributions of the characteristic states are obtained for all three stacking geometries (see Appendix~\ref{sec:appendix_boundary} and Fig.~\ref{efig_Boundary} for details).
For the alternated stacking, 
1D boundary states are found both between two nonpolar domains and at the interfaces between nonpolar and polar domains. 
In the helical stacking case, the three lowest-energy boundary states encircle the  hexagonal domains of Bernal stacking inside the hexagrams, forming a regular array of closed loops.
In contrast to the well-defined threefold rotational symmetry exhibited by the boundary states in the alternate and helical stacking cases, the boundary modes in the twisted monolayer-bilayer configuration [Fig.~\ref{efig_Boundary} (f)] exhibit nematic order, breaking the rotational symmetry. 
Importantly, unlike the boundary states in TTG, which hybridize with bulk metallic states~\cite{Park2025Nature}, the 1D boundary states in TTBNs are energetically well separated from the bulk states. 

Here, we highlight both the similarities and differences in the nature of quantum dot (QD) states in TTBN by comparison with other moiré systems. In twisted bilayer graphene (TBG), the electronic density becomes strongly localized near AA-stacking regions, primarily due to moiré-scale interference of interlayer tunneling processes rather than confinement by a real-space potential well. In this case, the relevant energy scale is set by the separation between the flat bands and the remote bands, typically on the order of ~30–60 meV, with a corresponding AA-centered localization length of ~5–10 nm~\cite{Bistritzer2011PNAS, Koshino2018PRX}. In contrast, in marginally twisted TMDC heterobilayers, localization arises from hydrostatic strain “hot spots” that form at the vertices of the reconstructed domain network. These strain fields generate confinement potentials of approximately ~100 meV, leading to QD bound states with characteristic sizes of ~2 nm~\cite{Enaldiev2022npj2D,Naik2018PRL}. For the TTBN structures studied here, the localization mechanism is instead governed by stacking-dependent electrostatic potentials that emerge at the vertices of the supermoiré domain network. Our calculations show that these potentials give rise to discrete bound QD states with substantially deeper confinement energies, ranging from ~200 to 400 meV, and characteristic localization lengths of about 1–2 nm.Overall, this comparison demonstrates that although vertex-localized states are a common feature across several moiré platforms, their microscopic origin and confinement strength differ significantly among TBG, TMDC heterobilayers, and TTBN.

\begin{figure*}[t]
\centering
\includegraphics[width=0.75\textwidth]{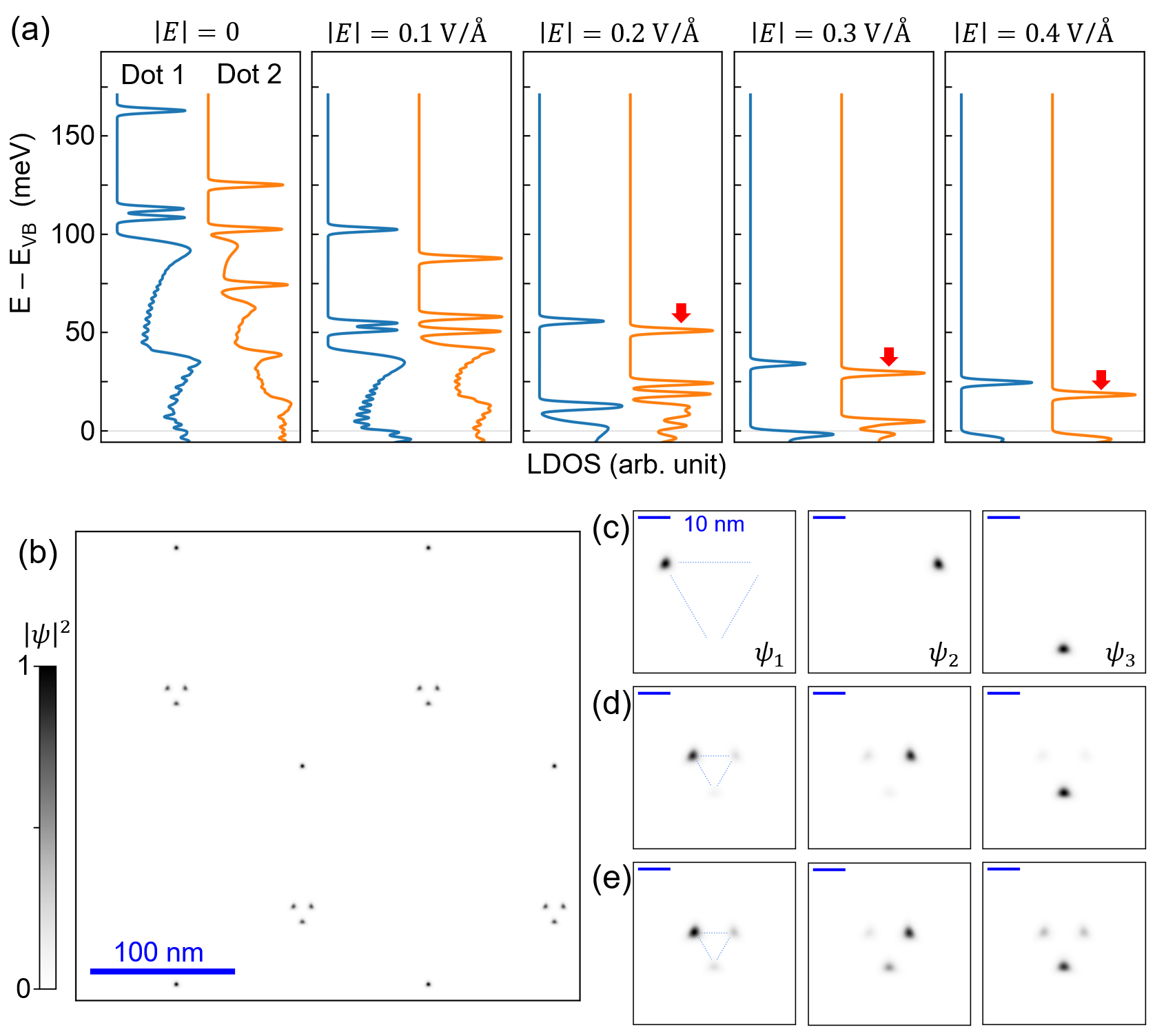}
\caption{
{Spatial and energetic variations of quantum dot eigenstates with electric fields.} 
(a) Variations of the local density of states (LDOS) at Dot 1 and Dot 2 of the alternate stacking case shown in Fig.~\ref{fig_alt_qho} (b) in the presence of the external electric field. From the left to the right panels, the electric field increases from 0.1  to 0.4 V/\AA~and corresponding relaxed structures are shown in Fig.~\ref{fig_domain_efield} (a). The valence band extrema from the center of domains for each field are set to zero. 
(b) Integrated spatial distribution of states above the bulk states energy of 0.0 eV under the electric field $\left|E\right|$ = 0.4 V/\AA. 
The modulus of three-fold degenerate wavefunctions for three Dot 2's forming equilateral triangles [their energies are denoted by red arrows in (a)], $\left|\psi_i\right|^2~(i=1,2,3)$ with the applied electric field, (c) $\left|E\right|$ = 0.2 V/\AA, (d) 0.3 V/\AA, and (e) 0.4 V/\AA. The colormap range is normalized to the maximum value in each panel.
}
\label{fig_QD_efield}
\end{figure*}

\subsection{Reconfigurable quantum dot arrays}

Having demonstrated that local domain shapes undergo significant changes under applied electric fields, we now explore the implications for the electronic structure of the localized states and their mutual interactions.
In Fig.~\ref{fig_QD_efield} (a), we show the local density of states (LDOS) for valence-band states at Dot 1 and Dot 2 in the alternate stacking configuration previously discussed in Fig.~\ref{fig_alt_qho}, as a function of the electric field. As the field increases, the confinement potential of each dot becomes shallower, and the energies of the lowest hole states approach the valence band edge of the bulk. Concurrently, the energy difference between the lowest hole states in Dot 1 and Dot 2 decreases and becomes nearly degenerate at a field strength of 0.3 V/\AA. This behavior is accompanied by a field-induced reduction in the band gap due to the Stark effect~\cite{Okada2011PRB}, which causes higher-energy localized hole states and boundary states to hybridize with bulk states, leaving only the lowest-energy localized hole states well isolated as demonstrated in Fig.~\ref{fig_QD_efield} (a).

As shown in Figs.~\ref{fig_domain_efield} (a) and (c), the interdot distance among the three Dot 2 sites—located at the vertices of the equilateral ACB-stacking triangles—decreases notably with increasing field. As a result, the spatial distribution of the localized hole states reveals a clustering effect, where three localized states move closer to each other to form equilateral triangles centered at the mass center of a larger triangle composed of Dot 1 sites as shown in Fig.~\ref{fig_QD_efield} (b).  
Simultaneously, interaction between localized states at Dot 1 and those at Dot 2 will be completely quenched with increasing fields. 

As the interdot spacing between Dot 2 sites decreases, their localized wavefunctions begin to overlap, leading to quantum hybridization. At each Dot 2 site, the lowest energy localized hole states, indicated by the red arrow in the middle panel of Fig.~\ref{fig_QD_efield} (a), are spin- and valley-degenerate.
For each dot site, we can assign the ground QHO state of $\varphi_i$ with $n_x=n_y=0$ for three sites ($i=1,2,3$), respectively.
Up to $E=0.2$ V/\AA, the states at the vertices of the triangle are well-localized,
so that each dot wavefunction can be written as $\psi_{\mathrm{D2}}^i = \varphi_i$ ($i=1,2,3$) as shown in Fig.~\ref{fig_QD_efield} (c). 
So, in this regime, we expect that the coupling between the triangular QDs can be well described by the capacitance model~\cite{Hsieh2012RPP,Chung2013PRB}. 
However, when the field increases above 0.3~V/\AA, the wavefunctions starts to hybridize, forming linear combinations of the localized states at each site, i.e., $\psi_{\mathrm{D2}}^i = c^i_1\varphi_1 + c_2^i\varphi_2 + c_3^i\varphi_3$ where coefficients of $c^i_j$ $(i,j=1,2,3)$ are set by detailed interactions between QDs. We found that the valley and spin-degenerate ground states are slightly split by about 0.1 meV while maintaining their coupling through tunnelings. For example, at the upper-left vertex of the triangle shown in the left panel of Fig.~\ref{fig_QD_efield} (d), the mixing ratios are $|c_2^1 / c_1^1|^2 = 0.24$ and $|c_3^1 / c_1^1|^2 = 0.12$. Similar mixing behavior is observed at the other two Dot 2 sites in the middle and right panels.
As shown in Fig.~\ref{fig_QD_efield} (e), with a further increase of the field, the hybridization becomes stronger as can be inferred from increasing mixing ratio of $|c_2^1 / c_1^1|^2 = 0.33$, and $|c_3^1 / c_1^1|^2 = 0.22$ for the same QDs.
This phenomenon is turned on at the critical field $E=0.22$ V/\AA, where the corresponding ratio is finite but marginal ($< 0.01$).

\begin{figure}[t]
\centering
\includegraphics[width=1.0\columnwidth]{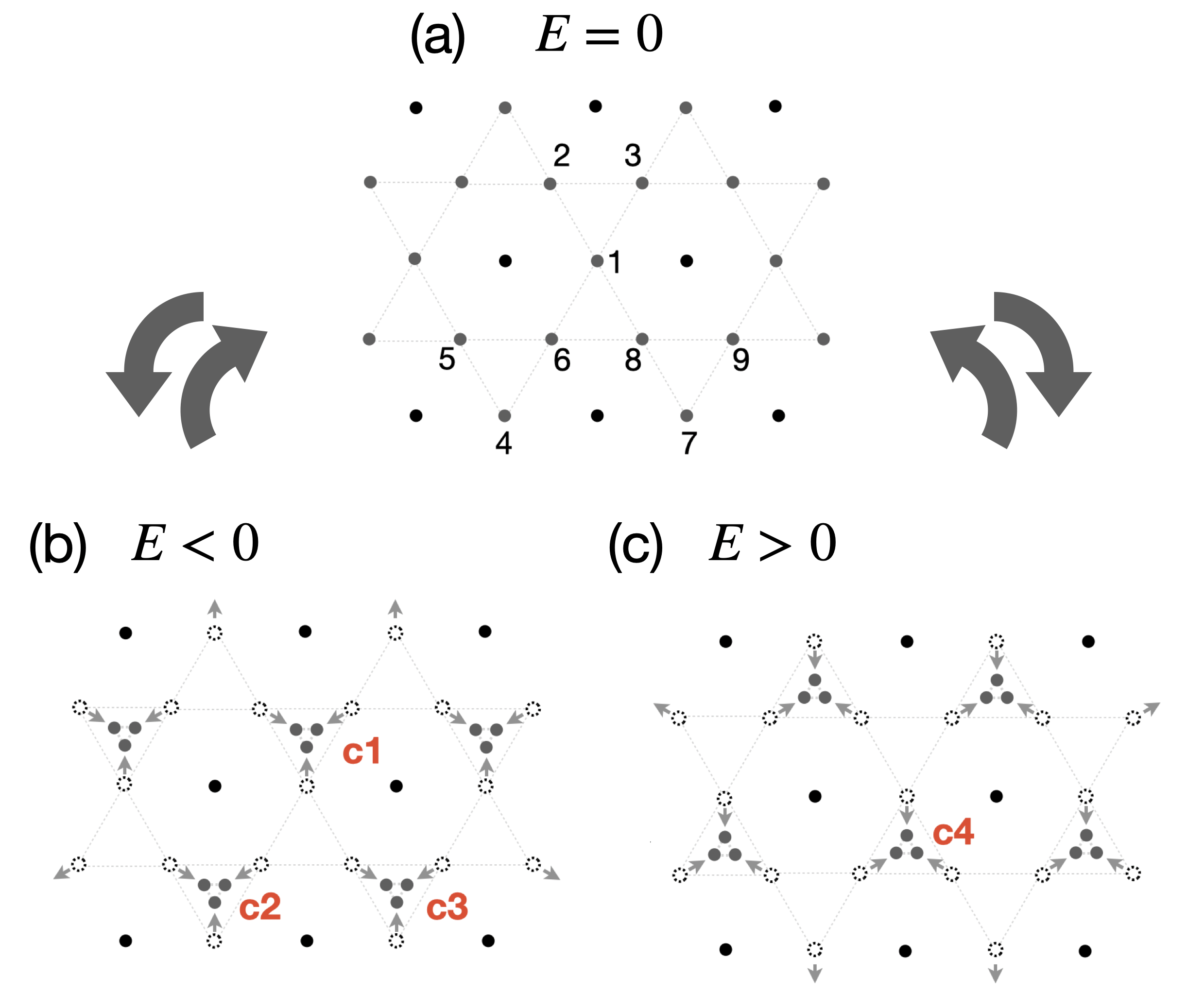}
\caption{
(a) Schematic redrawing of the localized quantum harmonic oscillator (QHO) states shown in Fig.~\ref{fig_alt_qho} associated with individual sites $i = 1, \dots, 9$ without electric field. (b) With an electric field ($E<0$ along polarization of ABC stacking), the reconfigured QD states form clusters denoted as c1, c2 and c3. Gray arrows indicate movement direction of each QD and empty dotted circles indicate their original positions. (c) With the opposite directional field ($E>0$), QDs form different clusters and one of them is denoted as c4. 
}
\label{efig_recluster}
\end{figure}

\section{Discussions and conclusions}

All the calculations hitherto confirm that the external electric field enables control over the spatial arrangement of QDs, resulting in clustering within certain regions while causing others to move apart. 
Under the field, the isolated localized states without the field begin to hybridize across neighboring QDs. 
When they form clusters, the systems we present exhibit other intriguing local features. 
The equilateral triangular configuration of three QDs shown in Fig.~\ref{fig_QD_efield} suggests potential applications in charge- and spin-based interactions, including spin qubits~\cite{Loss1998PRA}, due to several fascinating aspects such as spin or charge frustrations, topological properties and Nagaoka ferromagnetic molecular states, to name a few~\cite{Hsieh2012RPP,Chung2013PRL,Hamo2016Nature,Dehollain2020Nature}. 
To achieve these conditions, one must address the unavoidable fluctuations arising from the nuclear spins. Unlike silicon and germanium, which possess zero nuclear spins, B and N atoms cannot completely nullify their nuclear spins~\cite{Liu2022MQT}. However, recent advances in isotope engineering~\cite{Gong2024NatComm} and dynamical decoupling techniques~\cite{Rizzato2023NatComm} have demonstrated significant improvements in the coherence of spin defects in bulk h-BN materials. 
So, if properly prepared, the aforementioned intriguing charge- and spin-related physics can be realized in TTBNs with added features such as dynamic repositionings of QDs as discussed so far. 

We can envisage further a possible long-ranged information shuttling in solid state systems. 
For the alternate stacking geometry, if the applied electric field is first turned on along the polarization direction of ABC stacking, the domain area of ABC stacking increases while all the other domains shrink as already shown in Fig.~\ref{fig_domain_efield} (a). 
Then, if the field is reversed in direction, 
the enlarged ABC stacking area will return to its original shape and subsequently contracts, while the contracted ACB stacking area begins to expand. 
This can be expected well considering the upside down configuration of Fig.~\ref{fig_domain_efield} (a). 
During the process, field-induced coupled states in a triangular QD cluster first move apart as the field is reduced to zero.
Then, upon reversing the field direction, these states reassemble into new triangular clusters composed of QDs from different initial clusters, facilitating long-ranged transport of quantum states encoded in a coupled QD state across the array (see details in Fig.~\ref{efig_recluster} as was recently demonstrated in a Rydberg atoms array~\cite{Bluvstein2022Nature}.

Specifically, as illustrated in Fig.~\ref{efig_recluster}, the localized quantum harmonic oscillator (QHO) states, $\varphi_{i}$, can be associated with individual sites $i = 1, \dots, 9$ without electric field as denoted in Fig.~\ref{efig_recluster} (a). Upon application of an external electric field along the polarization direction of ABC stacking, the QD states corresponding to sites 1, 2 and 3 undergo spatial reorganization, forming a cluster of c1 denoted by red in Fig.~\ref{efig_recluster} (b). Concurrently, the states at sites 4, 5, 6 and 7, 8, 9 aggregate into distinct clusters, c2 and c3, respectively. Each of these cluster states, $\psi_{c_i}$ ($i = 1,2,3$), can realize either capacitive coupled or tunnel-coupled states. 
If the field is then adiabatically reduced to zero, the cluster states are now redistributed at sites 1, 6, and 8 in Fig.~\ref{efig_recluster} (a), respectively. Reversing the field to $E>0$ in Fig.~\ref{efig_recluster} (c) induces a second reconfiguration, in which these three states converge to form a new coupled state at the cluster c4. 
Backtracking the transformations from the right to left panels (changing electric field from positive to negative directions again), we can then return to the original situation. 
If coherence is maintained within local clusterings and regroupings, it becomes possible to observe the encoded coupled states that connect distant regions, spanning more than a few hundred nanometers, despite each individual dot moving back and forth locally.

We remark that the electric reconfiguration of QDs and their potential for the delivery of quantum coupling is not limited to the alternated stack TTBN.   
The helical stacking (hexagram domain pattern) and the twisted monolayer–bilayer stacking (kagome) TTBNs also host QD states at lower symmetric vertices that can realize the electric reconfigurations with different features and domain network topologies.
For example, in the helical stacking TTBN, a honeycomb array of QDs exhibits alternating pairings of QDs, an initial hexagonal unit into three pairs, under an electric field as shown in Fig.~\ref{efig_Helical_QHO} in Appendix~\ref{sec:appendix_QD_others}. 
In the kagome TTBN, alternating pairings in a linear array of QDs that also approach the 1D boundary states are expected, as shown in Fig.~\ref{efig_MonoBi_QHO} in Appendix~\ref{sec:appendix_QD_others}.

Lastly, as highlighted in previous studies on semiconducting twisted bilayer systems~\cite{Kennes2021NP,Enaldiev2022npj2D}, we also consider the significant potential that QD arrays offer for photonic applications, including single-photon emission (SPE)~\cite{Tran2016NatNano,Grose2020APLPh,Su2022NatMat}. h-BN is well established for such applications and demonstrates efficient deep ultraviolet emission and room-temperature SPE capabilities~\cite{Tran2016NatNano,Caldwell2019NatRevMat}. Given the excellent optical properties associated with defects in monolayer and bulk h-BN, our proposed system having arrays of localized QDs could play an important role in the realization of movable and programmable SPE arrays.
Many of the fundamental characteristics in h-BN and its twisted forms such as stacking geometries, domain polarization, and sliding ferroelectricity are also prevalent in transition metal dichalcogenides (TMDCs)~\cite{Andrei2021NatRevMat,Kennes2021NP}. 
So, we anticipate that certain theoretical predictions derived from our study may also be applicable to marginally twisted trilayer TMDCs.

In summary, we have shown that TTBN exhibits structurally and electronically tunable moir\'e patterns that arise from the interplay of multiple stacking and twisting degrees of freedom. These features distinguish TTBN from bilayer systems, where the domain structure is solely determined by a single twist angle. In TTBN, we identify regular arrays of localized quantum dot states at the vertices of super moir\'e domains, which can host well-defined localized quantum states with various spatial symmetries and nonzero angular momentum. The spatial arrangement and coupling strength of these states can be modulated by external electric fields, allowing for transitions between isolated and interacting regimes. This electric-field tunability, together with the well-defined domain structure and compatibility with large-scale fabrication, suggests that TTBN provides a versatile platform for investigating controllable quantum states and designing moir\'e-based quantum electronic devices.

\section*{Acknowledgments}
We gratefully acknowledge fruitful discussions with Mikito Koshino, Yun-Pil Shim and Gil Young Cho. 
K.Y. was supported by Korea Institute for Advanced Study (KIAS) individual grants (CG092501 and CG092502). 
C.P. was supported by the National Research Foundation of Korea (NRF) grant funded by the Korea government (MSIT and MOE) (No. RS-2025-16063688), and the Basic Science Research Program (No. RS-2024-00349573) through NRF, funded by the Ministry of Science and ICT of the Republic of Korea.
Y-W.S. was supported by the KIAS individual Grant No. (CG031509). Computations were supported by Center for Advanced Computation of KIAS.

\subsection*{Data availability}
All data is available in the manuscript.

\appendix

\section{Atomic Structure Optimizations}\label{sec:appendix_atom}

\begin{figure}[t]
\centering
\includegraphics[width=1.\columnwidth]{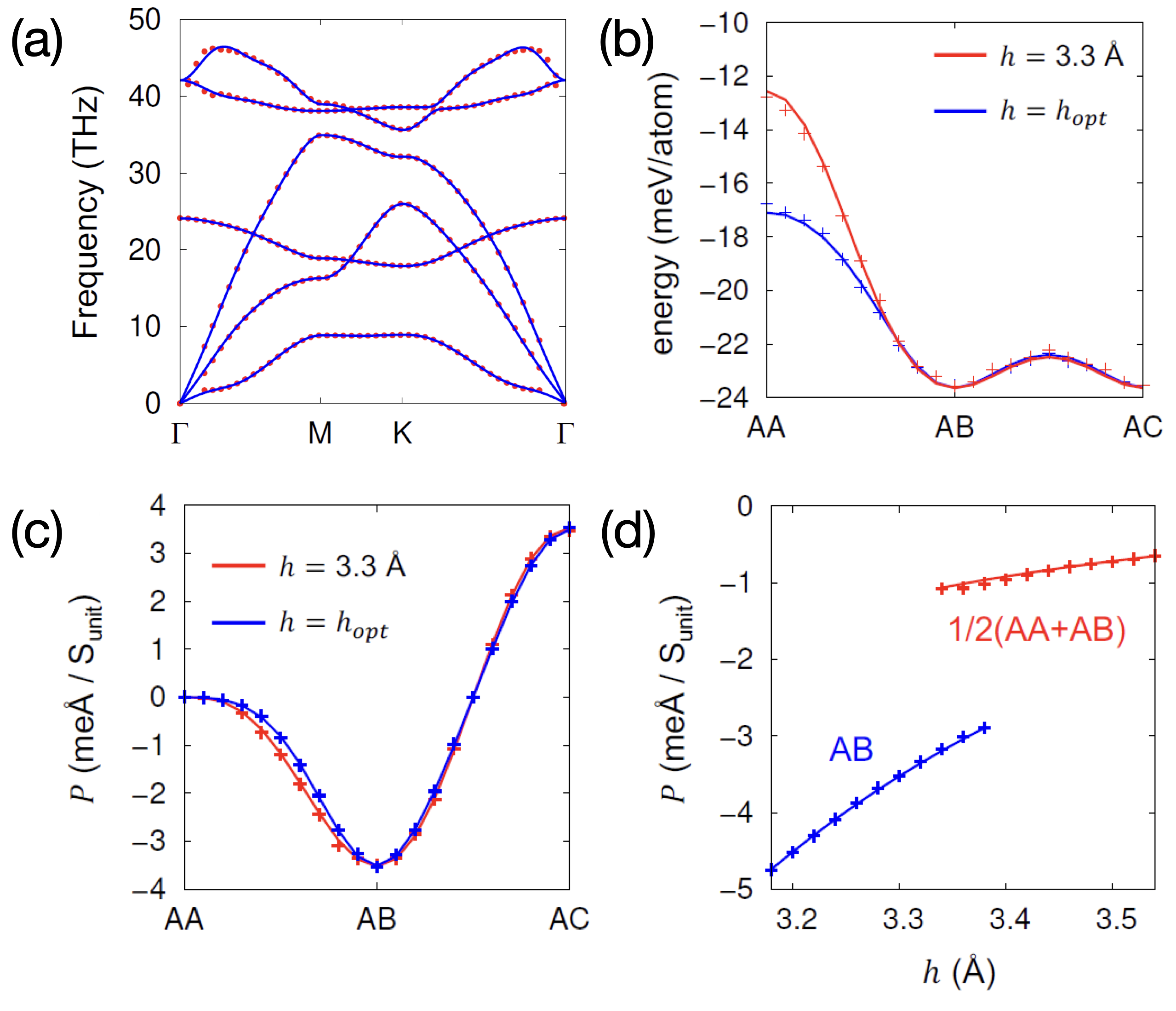}
\caption{{Interatomic potential of stacked hBN systems.} 
(a) Phonon dispersions of monolayer h-BN from first-principles calculations based on density function perturbation theory (dots) and our interatomic potential (lines). 
(b) Sliding energetics of parallel-stacked bilayer h-BN calculated from DFT-SCAN-rVV10 (dots) and our interatomic potential methods (lines) as a function of relative positions between two layers. 
Red (blue) denotes configurations in which the interlayer distance $h$ is fixed at 3.3~\text{\AA} (optimized individually) for each sliding configuration. 
(c) $P$ calculated using DFT-SCAN-rVV10 (dots) and our fitting function, Eq.~\eqref{eq:Pzfunc} (lines) with slidings.  (d) Variations of interlayer distance $h$ for AB and 1/2(AA+AB) stacking (see the text for definitions). Dots from the first-principles and lines from our method.
}
\label{efig_InterPot}
\end{figure}

\begin{table*}[t]
\caption{{Parameters for interlayer Kolmogorov-Crespi potential of multilayer boron nitrides.}}\label{etable_KC}
\centering
\begin{ruledtabular}
\begin{tabular}{lcccccccc}
  & $z_0$ (\AA) & $C_0$ (meV) & $C_2$ (meV) & $C_4$ (meV) & $C$ (meV) & $\delta$ (\AA) & $\lambda$ (\AA$^{-1}$) & $A$ (meV)\\
\colrule
B--B & 3.4451 & 15.1169 & 12.1952 & 4.9447 & 2.0358 & 0.9162 & 4.4219 & 12.8737 \\
B--N & 3.0024 & 14.8647 & 12.2137 & 4.9474 & 2.2505 & 1.0502 & 6.5692 & 13.4167 \\
N--N & 3.0260 & 15.2455 & 12.2318 & 4.9333 & 2.5001 & 0.4282 & 4.9484 & 12.0878 \\
\end{tabular}
\end{ruledtabular}
\end{table*}

We employed a newly developed interatomic potential method~\cite{Park2023NatComm,Park2025Nature} to compute atomic forces and obtain fully relaxed TTBN structures. 
This method is specifically designed to accurately replicate {\it ab initio} calculation results for large-scale systems, and effectively captures the delicate structural phase transitions in various materials. 
The harmonic intralayer potential is extracted from the phonon dispersion of monolayer boron nitride, which is obtained from our density functional theory (DFT) calculation with a generalized gradient approximation (GGA)~\cite{Perdew1996} using density functional perturbation theory (DFPT)~\cite{Baroni2001}. 
In Fig.~\ref{efig_InterPot} (a), phonon dispersions obtained by DFPT (dots) and our interatomic potential (lines), respectively, are compared.

For interlayer interactions, we incorporate additional pairwise forces by modifying the Kolmogorov-Crespi (KC) potential~\cite{Kolmogorov2005PRB,Park2025Nature} to account for long-range interlayer forces across all three layers~\cite{Park2025Nature}.
The original KC potential is given by the following form~\cite{Kolmogorov2005PRB,Naik2019}:
\begin{equation}\label{eq:KCform}
V_{ij} =e^{-\lambda(r_{ij}-z_0)}[C+2f(r_{ij,\parallel})]-A\left( \frac{r_{ij}}{z_0}\right)^{-6},
\end{equation}
where $r_{ij}$ and $r_{ij,\parallel}$ are the distance and in-plane distances between two atoms  $i$ and $j$ , respectively, and {$f(r)=e^{-(r/\delta)^2}\sum_{n=0}^{2}C_{2n}(r/\delta)^{2n}$}. 
The potential is defined for all pairs of atomic species, and each pairwise potential is constructed with eight parameters as presented in Table~\ref{etable_KC}. 
The parameters are fitted to reproduce the sliding energetics of bilayer BN calculated from DFT using SCAN+rVV10 for exchange-correlation and van der Waals functionals~\cite{Peng2016}.
For sliding configurations from AA to AB and to AC parallel stackings, the interlayer distance $h$ is either fixed to 3.3\AA~(red) or optimized (blue) as shown in Fig.~\ref{efig_InterPot} (b). 
Although the KC potential is quite accurate for bilayer systems~\cite{Kolmogorov2005PRB}, it cannot distinguish the energy of Bernal and rhombohedral stacking which is proved to be crucial in the small-angle ($<$ 0.1$^\circ$) TTG~\cite{Park2025Nature}. 
We also confirm this inaccuracy in multilayer h-BN. 
To correct this error, we added second nearest neighboring interlayer interactions to the original KC potential as was done for TTG~\cite{Park2025Nature}. 
A simple Gaussian-type potential $E_{2\mathrm{nn}} = -V e^{-3 r_{\parallel}^2}$ where
$r_{\parallel}$ is the layer-projected distance between the atoms in the first and third layers and $V=-0.4$ meV is used to make the energy difference between the ABA and ABC stacking h-BN to be $-0.42$ meV/atom, which is similar to the reported {\it ab initio} calculations~\cite{Haga2021,cortes2023}. 

\begin{table*}[t!]
\caption{{Parameters used in our tight binding models.} ``B-N 1st'' indicates the nearest neighboring B and N atoms on intralayer or B and N atoms between the interfacing two layers (interlayer). Others follow the same rule. }\label{etable_TBparam}
\centering
\begin{ruledtabular}
\begin{tabular}{lccc|ccccc}
\multicolumn{4}{c|}{Intralayer hopping parameters}&
\multicolumn{5}{c}{Interlayer hopping parameters}\\
\colrule
Bond type  &  $d_0$ (\AA) &  $t_0$ (eV) &  $s$ (eV/\AA) &  $V^0_{\sigma}$ (eV) &  $\beta_{\sigma}$  &  $V^0_{\pi}$ (eV)  &  $k_{\pi}$ (\AA$^{-1}$)  &  $d_{\pi}$ (\AA)\\
\colrule
B--N 1st & 1.4457 & $-2.6909$ & 5.1067 & 0.3233 & 4.4827 &$ -0.4953$ & 1.5533 & 4.3920 \\
B--B 1st & 2.5040 & $-0.2449$ & 1.5794 & 0.4086 & 4.3051 & $-1.2246$ & 1.5254 & 4.0241 \\
N--N 1st & 2.5040 & $-0.2138$ & $-0.6285$ & 0.3802 & 3.2378 & $-0.4145$ & 2.2434 & 4.0699\\
\colrule
B--N 2nd & 2.8914 & 0.0321 & 0.1510 & \multicolumn{5}{c}{See text for 2nd (NNL interaction)}\\
B--B 2nd & 4.3371 & $-0.1170$ & 0.0307 \\
N--N 2nd & 4.3371 & $-0.0583$ & 0.1026 \\
B--N 3rd & 3.8249 & 0.2500 & $-0.2743$ \\
B--B 3rd & 5.0080 & $-0.1019$ & $-0.0084$ \\
N--N 3rd & 5.0080 & $-0.0798$ & 0.1464 \\
B--N 4th & 5.2125 & 0.0309 & $-0.0393$ \\
\end{tabular}
\end{ruledtabular}
\end{table*}

Between the two layers of h-BN, there can be finite out-of-plane electric polarization of $P_{z}$, which determines the response to the external electric field. 
For example, for AC (or BA) stacking, $P_{z}>0$ develops. 
For 180$^\circ$-rotated bilayer, by symmetry, $P_{z}=0$. 
$P_{z}$ between two layers are described by our pairwise dipole model,
\begin{equation}\label{eq:Pzfunc}
P_z(\mathbf{r}_{\text{BN}}) = P_0\frac{z}{r}
e^{-\lambda(r-z_0)}e^{-kr_{\parallel}^2},
\end{equation}
where $\mathbf{r}_{\text{BN}}$ is a vector from a B atom to a N atom between two neighboring layers, and $r$, $r_{\parallel}$ and $z$ are $\left|\mathbf{r}_{\text{BN}}\right|$, in-plane and $z$-component of $\mathbf{r}_{\text{BN}}$, respectively.
$P_0$, $\lambda$, $z_{0}$ and $k$ are fitting parameters.
We found that the screening-like term $e^{-\lambda(r-z_0)}$ alone cannot accurately reproduce the reference calculation results. So, the Gaussian overlap-type fitting function $e^{-kr_{\parallel}^2}$ is additionally introduced.  
The total polarization of the system is $P=\sum_{\{\mathbf{r}_{\text{BN}}\}}P_z(\mathbf{r}_{\text{BN}})$, where $\{\mathbf{r}_{\text{BN}}\}$ is the set of interlayer B-N vectors.
For our SCAN+rVV10 calculations for BBN, $h=3.3$ \AA~and $P_{z}= 4.50 \times 10^{-3}$ $e$\AA/A$_\text{unit}$
where A$_\text{unit}$ ($=5.43$ \AA$^2$) is a unit area. 
The four parameters are fitted to reproduce $P_{z}$ of sliding configurations in Fig.~\ref{efig_InterPot} (c) and additionally the $h$-dependence of $P_{z}$ for AB stacking and halfway between AA and AB stacking denoted as 1/2(AA+AB). 
The parameters $P_0 = 1.61\times 10^{-2}\ e$\AA, $\lambda = 1.94$ \AA$^{-1}$, $z_0 = 3.30$ \AA, and $k=2.17\times 10^{-1}$ \AA$^{-2}$ reproduce {\it ab initio} results for both configurations as shown in Figs.~\ref{efig_InterPot} (c) and (d).

\section{Electronic Structure Calculation Methods}\label{sec:appendix_electron}

To calculate electronic structures of TTBNs with multimillion atoms, we developed an efficient calculation method with tight-binding (TB) approximations.
Independent DFT calculations were performed as references of the electronic structures of the few-layer h-BN.
We adopted the self-consistent extended Hubbard corrected DFT method (DFT+$U$+$V$)~\cite{Lee_UV_2020PRR,Rubio2020PRB,Timrov2020prb,Yang2021PRB} implemented in our in-house version~\cite{inhouse} of the Quantum ESPRESSO suite~\cite{QE1_2009, QE2_2017}.
DFT+$U$+$V$ method can efficiently reproduce the electronic band gap in an accuracy comparable to the $GW$ approximations without serious computational resources~\cite{Lee_UV_2020PRR,Rubio2020PRB,Timrov2020prb}. 
We calculated the self-consistent on-site ($U$) and intersite ($V$) Hubbard interactions for the monolayer h-BN
and used them for the bilayer and trilayer systems. We checked the self-consistently determined $U$ and $V$ values in the different bi- and trilayer systems and confirmed no significant deviation from the monolayer case.
Obtained interactions are $U_{\text{B}p}$ = 0.5112 eV and $U_{\text{N}p}$ = 4.9670 eV, where the subscripts refer to the atom and orbital species, e.g., B$p$ indicate $p$ orbitals of B atom, and intersite parameters are $V_{\text{B}s-\text{N}s}$ = 1.0918 eV, $V_{\text{B}s-\text{N}p}$ = 1.9254 eV, $V_{\text{B}p-\text{N}s}$ = 1.4700 eV, and $V_{\text{B}p-\text{N}p}$ = 2.9767 eV. 
The energy cutoffs are 100 and 1000 Ry for wavefunction and charge density, respectively. The $k$-points grid is $\Gamma$-centered $24\times24\times1$. 
The lattice parameter is 2.504 \AA, the interlayer spacing $c_0=3.253$ \AA\, and the $z$-direction length of the unit cell including the vacuum 30 \AA. We also considered the dipole correction to the electrostatic potential.
Finally, we obtained Wannier functions constructed from low energy states by using Wannier90 code~\cite{Pizzi_wannier90_2020}. We adopted $p_z$ orbitals for each B and N atom as initial projectors. 

We constructed a TB model to investigate the electronic structure of the trilayer h-BN, which adopted the $p_z$ orbital at each atom as the basis. The hopping parameters are parametrized by fitting to the Wannier Hamiltonian obtained from the DFT+$U$+$V$ calculations.
The on-site orbital levels are given as 2.625 eV for B and $-3.495$ eV for N.
Since the deformation of each layer is smooth and moderately small, intralayer bond lengths also only slightly deviate from their unstrained references. So, for the intralayer hopping parameters, the Wannier Hamiltonian components of the corresponding bonds are adopted similarly to a previous study~\cite{Li2024PRB, javvaji_ab_2025}. Specifically, we considered only the intralayer bonds within the cutoff of 5.5 \AA, which correspond to up to the fourth nearest B-N bond and the third nearest of each B-B and N-N bond. To incorporate the strain effect, the intralayer hoppoing parameters vary linearly with respect to the bond length change from the reference structure, i.e., $t(r) = t_0 + s(r-d_0)$, where $r$ is the bond length, $t_0$ and $d_0$ are the reference hopping parameter and bond length of the unstrained system, respectively, and $s$ is the gradient of the hopping parameter. The gradient is determined from the Wannier Hamiltonians of the $\pm2\%$ strained systems. The detailed values of $d_0$, $t_0$, and $s$ for each type of bond are listed in Table.~\ref{etable_TBparam}.

For our TB model to reproduce all these reference data, the interlayer hopping is given by Slater-Koster type parametrization with decaying functions of distance, $t(\mathbf{r}) = n^2V_{\sigma}(r) + (1-n^2)V_{\pi}(r)$, where $n = z/\left|\mathbf{r}\right|$, $V_{\sigma}(r) = V_{\sigma}^0 \exp\left[-\beta_{\sigma}(r-c_0)/c_0\right]$, and $V_{\pi}(r) = V_{\pi}^0 \left[1+\tanh{(-k_{\pi}(r-d_{\pi}))}\right]/2$. 
The fitting references were Wannier Hamiltonian of AA, AB, and slightly displaced stackings from AB stacked bilayers. The cutoff distance of the interlayer hopping is 6 \AA.  The detailed parameters are listed in Table~\ref{etable_TBparam}.

To reproduce the degeneracy splitting in {\it ab inito} bands of the trilayer BN, we also considered next-neighboring layer (NNL) interactions. 
The NNL hopping parameters are given as $t^{\text{NNL}}(r) = t_0 \left[1+\tanh{\left(-\gamma(r-r_0) \right)}\right]/2$, where $t_0$ = 0.02, $-0.03$, and $-0.01$ eV for B-N, B-B, and N-N hoppings, respectively, $\gamma=5.96$ \AA$^{-1}$~and $r_0=6.753$ \AA. The cutoff distance of the NNL hopping is 7\ \AA.

\begin{figure*}
\centering
\includegraphics[width=0.75\textwidth]{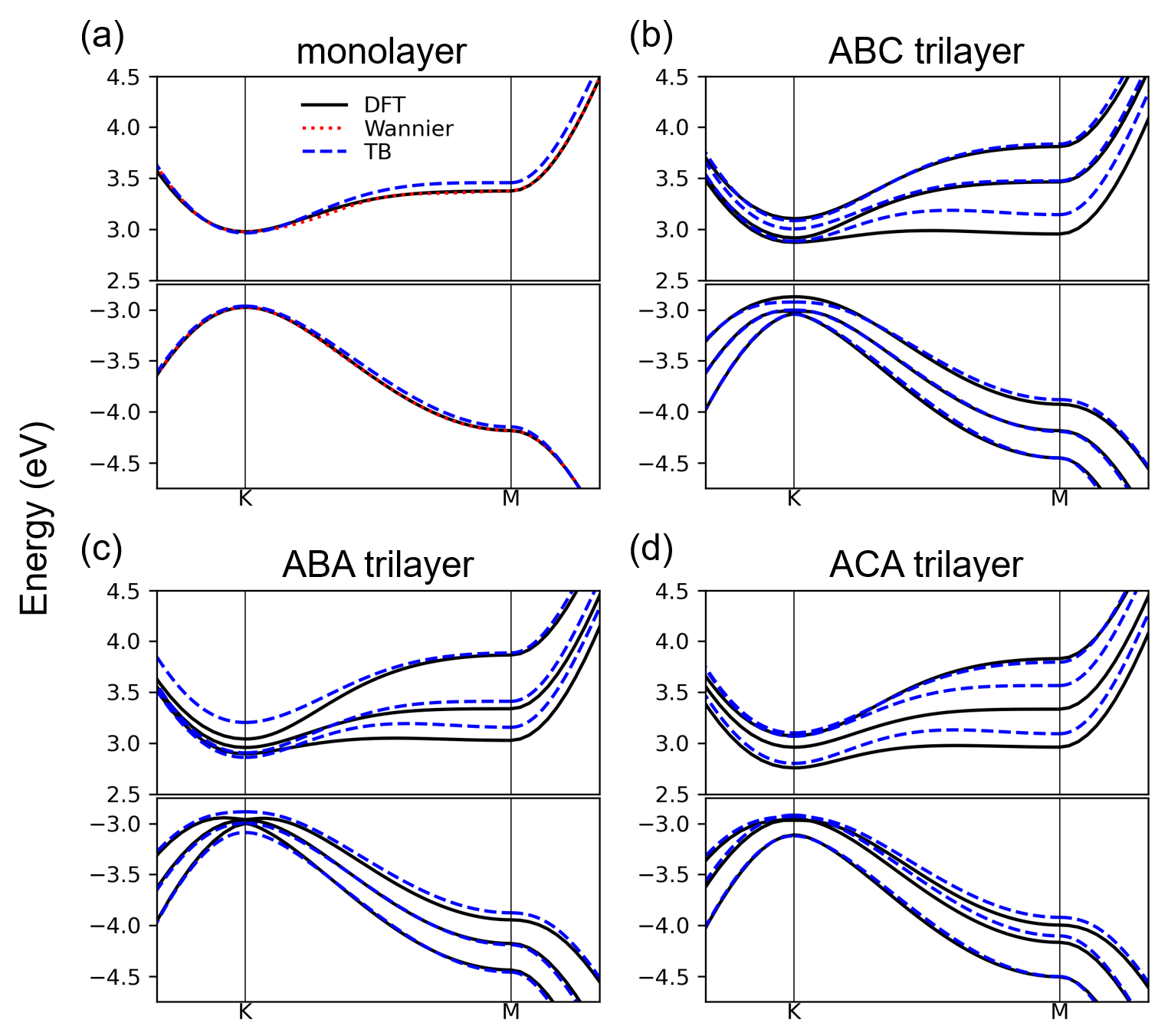}
\caption{
{Band structures near the CBM and VBM.}
(a) Monolayer h-BN band structures from DFT, Wannier Hamiltonian, and TB model calculations. Trilayer band structures from DFT and TB model calculations for (b) ABC, (c) ABA, and (d) ACA stackings.
}
\label{efig_TBbands}
\end{figure*}

In addition, we considered the on-site level correction that mimics the potential energy gradient induced by the local dipole moment. To calculate the local gradient, we first determined the atomic portion of the dipole moment $P_i$ of the atom $i$ for the interfaces between adjacent layers.
The preliminary atomic portion $P^{\text{pre}}_i$ is determined by dividing the dipole moment given by Eq.~\eqref{eq:Pzfunc} in half and assigning it to the two atoms determining $\mathbf{r}_{\text{BN}}$, and adding it up for every B-N pair. 
Then, $P_i$ is given as $P_i = P^{\text{pre}}_i/2 + \sum_{j\in \text{NN}} P^{\text{pre}}_j/6$, where NN means the three nearest neighbor atoms. In this way, we can determine the smoothly distributed atomic portion of the dipole moments preserving their total moment.  
The on-site correction term is given as $\Delta E_{\text{B/N}} = \mp c_{\text{B/N}}P_i$ for B and N atoms and $c_{\text{B}} = 82$ V/\AA, and $c_{\text{N}} = 40$ V/\AA, where the signs are for the upper and lower side of the interface, respectively.
The resulting TB band structures are shown in Fig.~\ref{efig_TBbands}.
The applied electric field is also simulated as the on-site energy level shift depending on the $z$-coordinate by $\Delta U(z) = zE_z/\epsilon_r$ with the fitted dielectric constant $\epsilon_r=1/0.413$ to reproduce the DFT results.

The moir\'e supercell contains about three million atoms, which is equivalent to the dimension of the Hamiltonian matrix. Since the full-diagonalization  is impractical, we only consider the low-energy eigenstates by using the ARPACK eigenvalue-solver which adopts the implicitly restarted Lanczos method~\cite{lehoucq_arpack_1998, opencollab_arpack-ng}. The linear equation solver implemented in the PETSc library is utilized for the shift-invert mode of the Lanczos scheme~\cite{petsc_1997}.

\section{Area Variation of Polarized Domains under Electric fields}\label{appendix:slideFE}

\begin{figure*}
\centering
\includegraphics[width=0.75\textwidth]{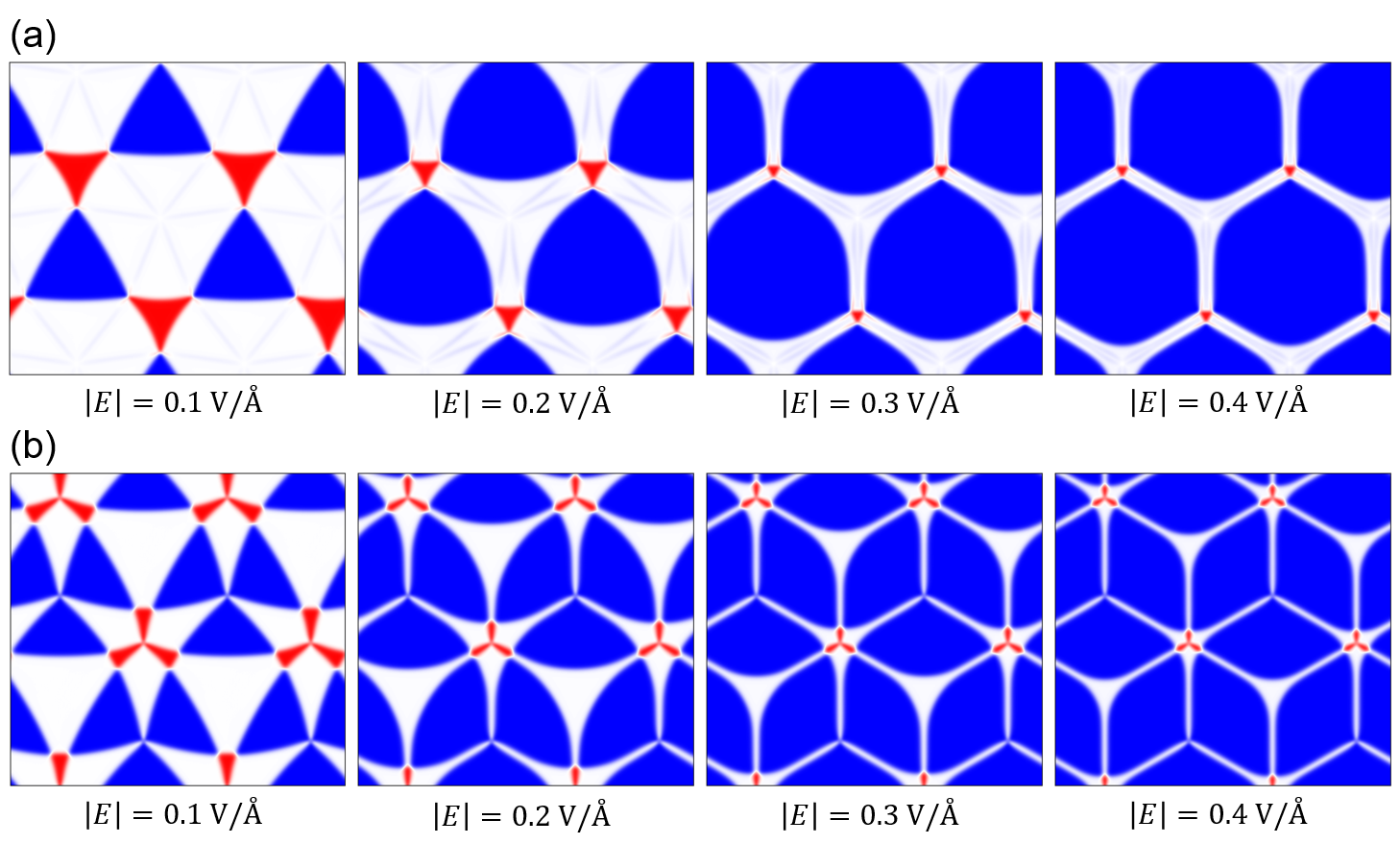}
\caption{
{Variation of polarization distributions of TTBN under electric fields.} 
(a) Changes of polarized domains of alternate stacking TTBN in Fig.~\ref{fig_domain_efield} (a) under a perpendicular electric field of which direction is parallel to the polarization of ABC (blue) stacking. 
(b) Same figures for helical stacking TTBN in Fig.~\ref{fig_domain_efield} (b).
}
\label{efig_Pol_Efield}
\end{figure*}

In this appendix, we discuss the field-induced variations in an area of polarized domains. 
In alternate and helical stackings shown in Fig.~\ref{fig_domain_efield} (a) and (b) respectively, 
with the external electric field parallel to the polarization direction of ABC stacking (blue in Fig.~\ref{efig_Pol_Efield}), the total energy of the ABC stacking decreases and then maximize their area while the nonpolar (white) and opposite polarization (red) area decrease simultaneously.

\begin{figure*}
\centering
\includegraphics[width=0.75\textwidth]{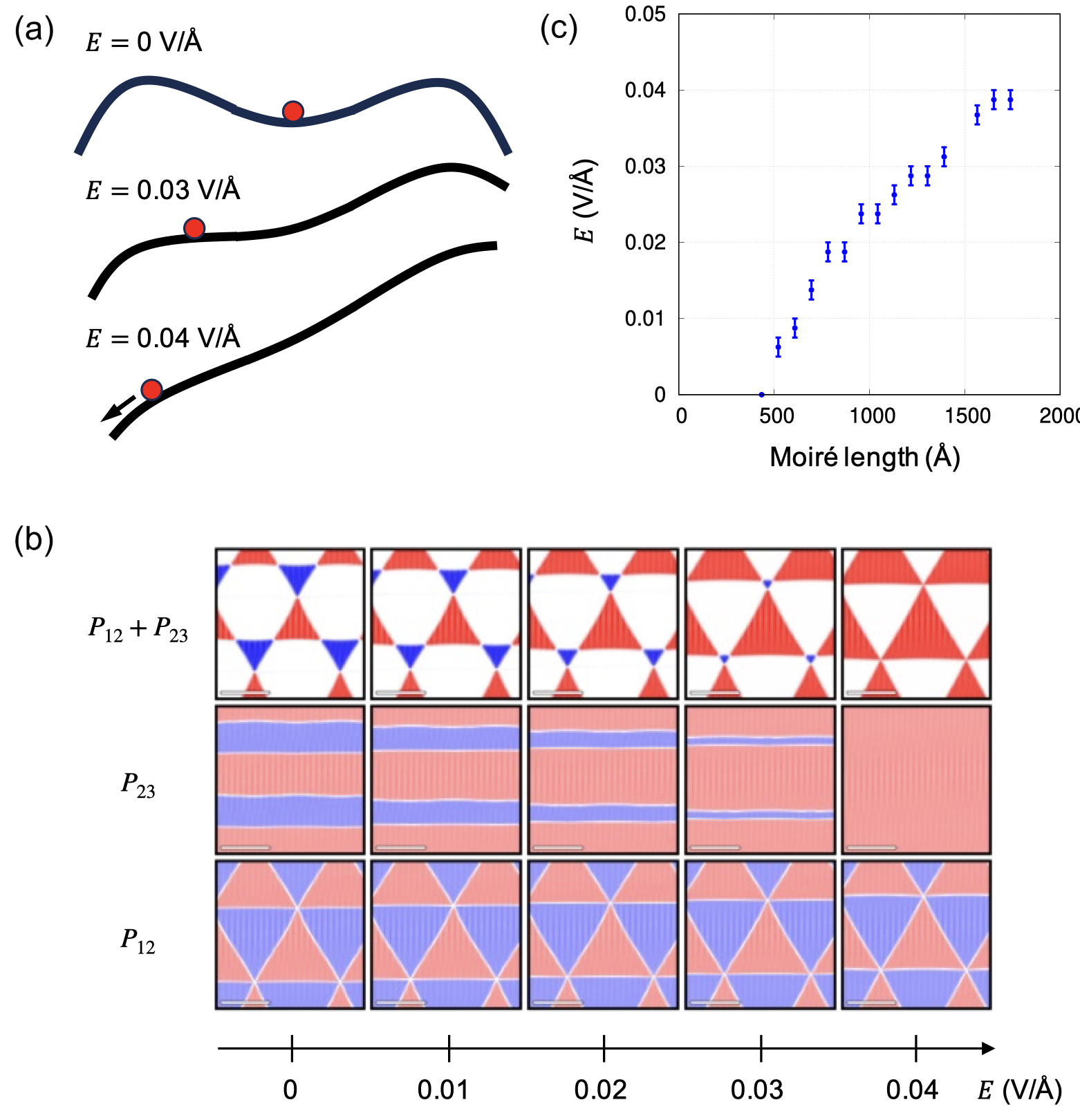}
\caption{
{Switching between kagome-shaped and triangular domains under electric fields}.
(a) Schematic description demonstrating the evolution of energy landscapes for $\theta_{12}=0.082^\circ$ when a small electric field $E$ is applied.
(b) Layer-resolved polarization densities ($P_{12}$ for the first and second layer and $P_{23}$ for the second and third one) as a function of applied electric field of $E$. Top panels are total polarization maps. 
(c) The critical fields for switching between a kagome domain lattice and a triangular domain lattice. As the moir\'e length $L$ decreases, the critical field also decrease and finally the kagome domain becomes unstable for $L<500$~\AA.   
}
\label{efig_Field_MonoBi}
\end{figure*}

\FloatBarrier

\begin{figure*}[t]
\centering
\includegraphics[width=0.9\textwidth]{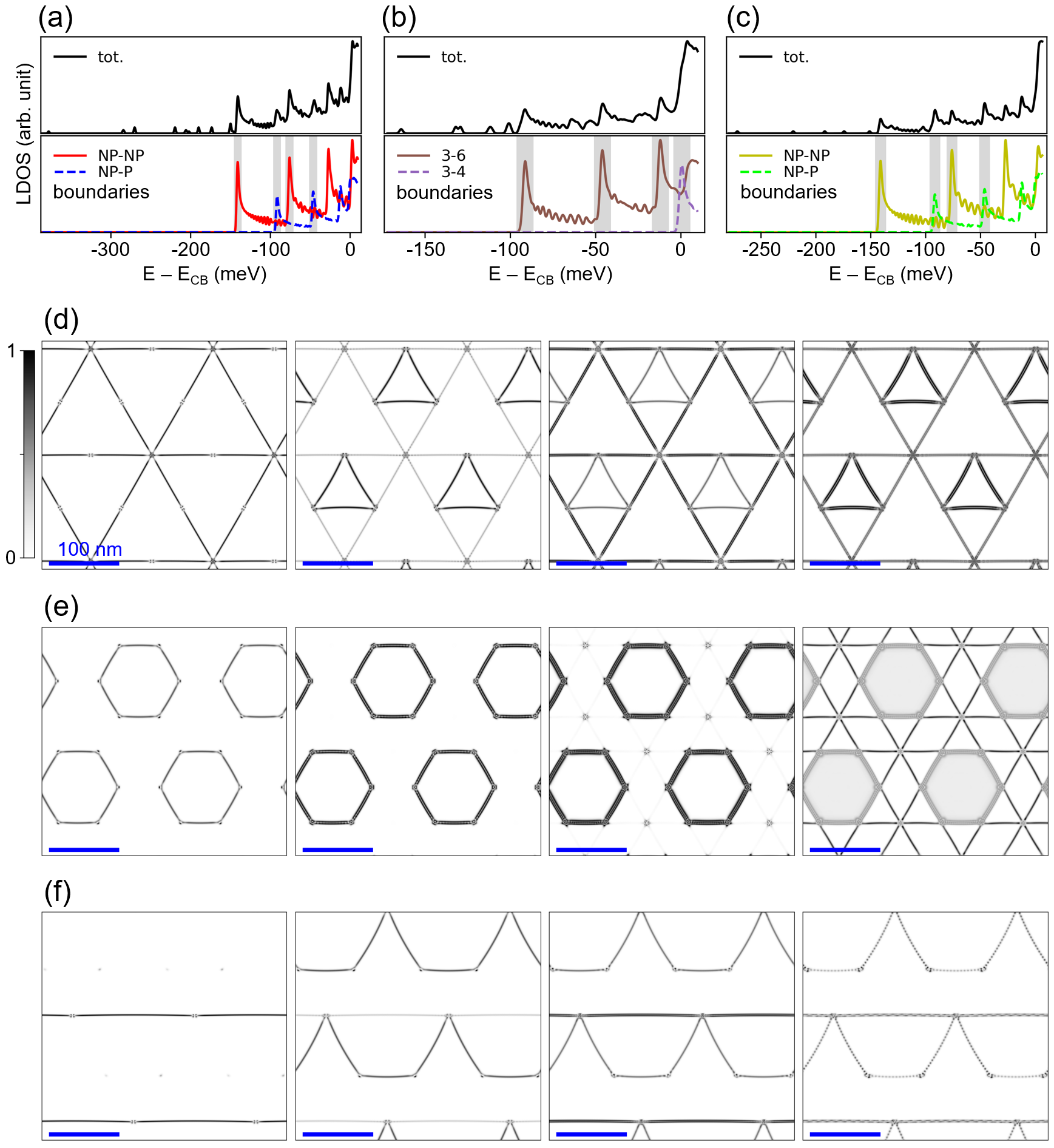}
\caption{
{Domain boundary states of TTBNs.}  
Shown are the total density of states (DOS; upper panels) and boundary-projected local density of states (LDOS; lower panels) for (a) alternate stacking, (b) helical stacking, and (c) twisted mono–bilayer TTBNs. In (a) and (c), “NP–NP” and “NP–P” denote LDOS at boundaries between nonpolar domains, and between nonpolar and polar domains, respectively. In (b), “3–6” and “3–4” refer to boundaries between triangular and hexagonal, and triangular and tetragonal domains.
Panels (d)–(f) display Gaussian-weighted spatial distributions of eigenstates for the corresponding stackings, centered at the LDOS peak energies. 
Each set of four panels is arranged in order of increasing energy.
}
\label{efig_Boundary}
\end{figure*}

In the twisted monolayer-bilayer system, a finite net polarization density modifies the potential energy landscape [illustrated in Fig.~\ref{efig_Sliding} (a) and (b)] under an external electric field, $E$ as $-P(\Delta x)\cdot E-\chi E^2$. 
The second term is negligible for small fields $E<0.1$ V/\AA. 
Figure~\ref{efig_Field_MonoBi}(a) shows the altered energy profile for a twist angle $\theta_{12}=0.082^\circ$ under a weak applied field.
Due to the shallow energy landscape around the kagome domain configuration, increasing $E$ eventually destabilizes this local minimum at a critical field strength of approximately $E=0.04$ V/\AA.  
Figure~\ref{efig_Field_MonoBi} (b) illustrates how the polarization domains in each layer evolve as a function of the electric field.
The dominant variation occurs in $P_{23}$, indicating that structural relaxation primarily involves the displacement of one-dimensional dislocation domain walls. 
The critical fields at which the kagome domain becomes unstable are plotted in Fig.~\ref{efig_Field_MonoBi} (c) with an uncertainty of 0.0025 V/\AA. 
As the moir\'e length $L$ decreases, the energy barrier is reduced as shown in Fig.~\ref{efig_Sliding} (a),  and the critical field vanishes for $L<500$ \AA, indicating that the kagome domain becomes unstable.

\begin{figure*}[t]
\centering
\includegraphics[width=1.0\textwidth]{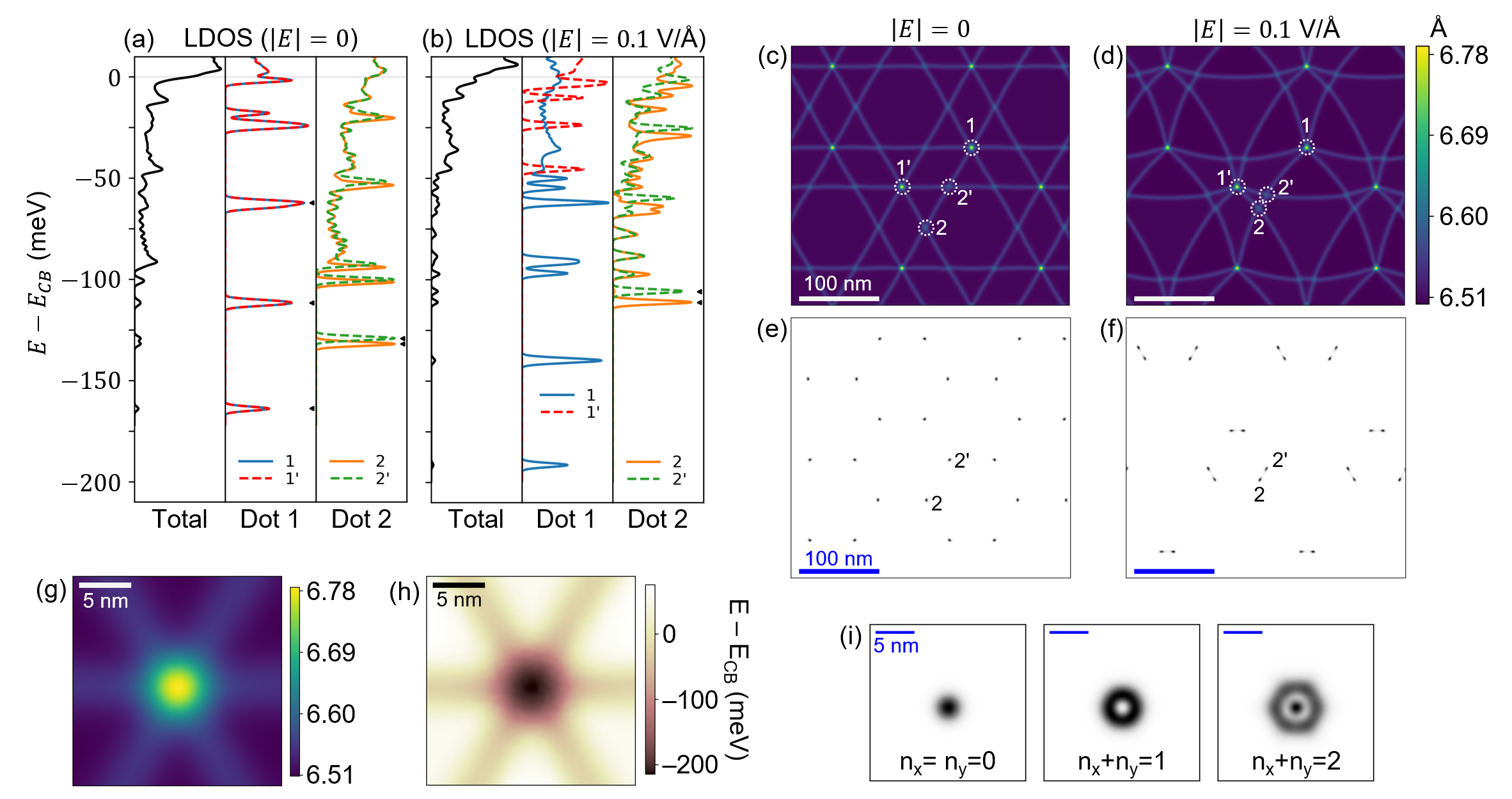}
\caption{
{QD states in the helical TTBN  ($\theta_{12}=\theta_{23}=\theta=0.082^\circ$) and their electrical reconfiguration.} 
(a) and (b) The total density of states (Total) and local density of states (LDOS) of conduction bands for (a) $\left|E\right|=0$ and (d) $\left|E\right|=0.1$ V/\AA, where the projections are for sites 1, $1'$, 2, and $2'$ in (c) and (d), respectively. The energy level is a relative value to the conduction edges $E_\text{CB}$ of the bulk trilayer stackings. 
(c) and (d) The fully relaxed interlayer spacing between the bottom and top layers drawn for the same region as Fig.~\ref{fig_domain} (d) and the first panel of Fig.~\ref{fig_domain_efield} (b), respectively.
(e) The density distribution of the lowest QD states at sites 2 and $2'$ shown for the same region as (c), of which energies are indicated by small black triangles in (a) Dot 2 LDOS.
(f) Similar data but under the electric field $\left|E\right|=0.1$ V/\AA, for the same region as (d) and energy peaks indicated in (b) Dot 2 LDOS.
(g) The enlarged interlayer spacing around the site 1 in (c). (h) The potential well of the conduction band energy landscape at QD site 1 shown in (g). 
(i) The enlarged density distributions of the first three lowest QD site 1 states, of which energies are indicated in (a) Dot 1 LDOS. For each panel, corresponding quantum numbers of isotropic 2D QHO states $n_x+n_y$ are displayed. 
} 
\label{efig_Helical_QHO}
\end{figure*}

\section{One-dimensional Domain Boundary States}\label{sec:appendix_boundary}

We present a detailed analysis of domain boundary states in various TTBN stacking geometries by comparing the total density of states (DOS) and the boundary-projected local density of states (LDOS) for the three stacking configurations shown in Fig.~\ref{fig_domain}. In the case of alternate and twisted mono–bilayer stackings shown in Figs.~\ref{efig_Boundary} (a) and (c), the boundary-projected LDOS is further resolved according to the polarization nature of the adjacent domains. For the boundaries between two nonpolar domains, the well-defined 1D boundary DOS appears. Likewise, the LDOS between polar and nonpolar domains also supports 1D states. For the helical stacking geometry illustrated in Fig.~\ref{efig_Boundary} (b), domain boundaries are labeled according to their adjacent domain types—specifically triangular,  hexagonal, and tetragonal domains—highlighting the geometric complexity of corner-shared hexagram domain tessellations unique to this configuration.

To further elucidate the spatial localization and energy characteristics of these domain boundary states, we calculate Gaussian-weighted integrals of the eigenstate distributions centered at representative boundary LDOS peak energies. These are shown in panels Figs.~\ref{efig_Boundary} (d) - (f) for alternate, helical, and twisted mono–bilayer stackings, respectively. The Gaussian weight function is centered at the LDOS peaks identified in panels Figs.~\ref{efig_Boundary} (a)–(c), with a smearing width of $\sigma = 5$ meV; the corresponding $\pm\sigma$ energy windows are indicated by gray-shaded regions in the LDOS plots. For clarity and systematic comparison, the four subpanels in each of Figs.~\ref{efig_Boundary} (d)–(f) are arranged in ascending order of energy, allowing for an understanding of how the electronic states evolve across the energy spectrum in relation to their localization at domain boundaries. 
We note that all the boundary states are well localized at the interfaces between domains except a boundary state with an energy around 0 meV for the helical stacking in Fig.~\ref{efig_Boundary} (b). This state is a mixture of extended states on hexagonal domains and localized boundary states as shown in the fourth panel of Fig.~\ref{efig_Boundary} (e).

\begin{figure*}[t]
\centering
\includegraphics[width=1.0\textwidth]{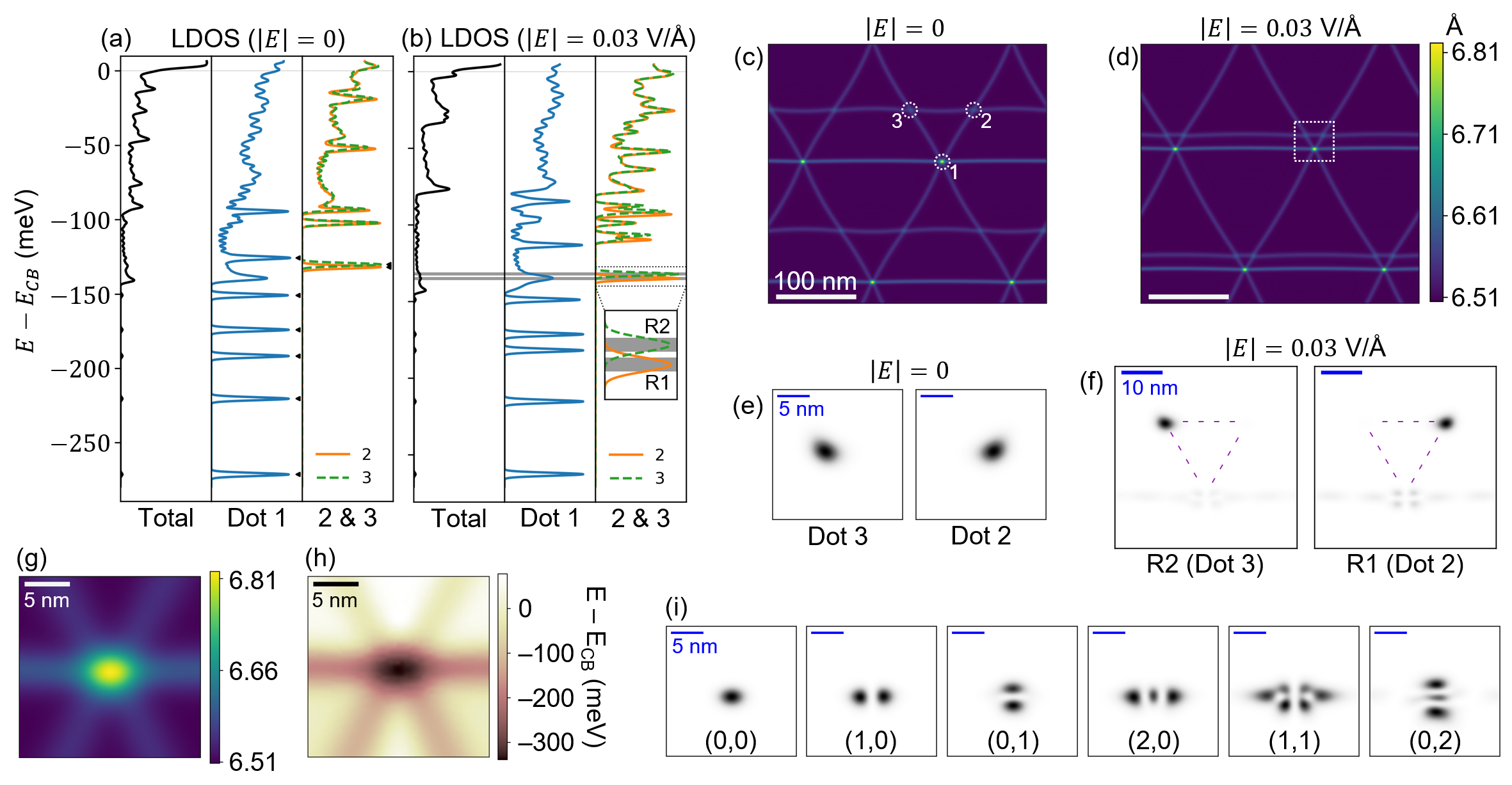}
\caption{
{QD states in the twisted monolayer-bilayer TTBN ($\theta=0.082^\circ$) and their electrical reconfiguration.}
(a) and (b) The total density of states (Total) and local density of states (LDOS) of conduction bands for (a) $\left|E\right|=0$ and (d) $\left|E\right|=0.03$ V/\AA, where the projections are for sites 1, 2, and 3 indicated in (c) and corresponding sites in (d), respectively. The energy level is a relative value to the conduction band edges $E_\text{CB}$ of the bulk trilayer stackings. 
(c) and (d) The fully relaxed interlayer spacing between the bottom and top layers drawn for the same region as Fig.~\ref{fig_domain} (f) and equivalent area under the field $\left|E\right|=0.03$ V/\AA, respectively.
(e) The enlarged density distribution of the lowest QD states at sites 2 and 3, the energies of which are indicated by small black triangles in (a) Dot 2 LDOS.
(f) For the square area marked in (d), Gaussian-weighted spatial distributions of eigenstates under the electric field $\left|E\right|=0.03$ V/\AA\ corresponding to range 1 (R1) and range 2 (R2) indicated by shadings in (b). The shadings show the $\pm\sigma$ (broadening factor, 1 meV) range around the Gaussian-weight center.
(g) The enlarged interlayer spacing around the site 1 in (c). (h) The potential well of the conduction band energy landscape at QD site 1 shown in (g). 
(i) The enlarged density distributions of the first six lowest QD states at site 1, of which energies are indicated in (a) Dot 1 LDOS. For each panel, corresponding quantum numbers of 2D QHO states $(n_x,n_y)$ are displayed. 
} 
\label{efig_MonoBi_QHO}
\end{figure*}

\section{Quantum Dot States in Helical Stacking and Twisted monolayer-bilayer Systems}\label{sec:appendix_QD_others}

We investigate the relation between structural and electronic characteristics of QD states in TTBN with both helical and monolayer–bilayer configurations, as well as their electric reconfigurations.
For the helical TTBN, Fig.~\ref{efig_Helical_QHO} (c) shows the spatial variation of the fully relaxed interlayer spacing between the bottom and top layers without an applied field for the same region as Fig.~\ref{fig_domain} (d). 
Therein, localized increases in the interlayer distance imply QD sites labeled as 1, $1'$, 2, and $2'$. 
The enlarged view of site 1 is in Fig.~\ref{efig_Helical_QHO} (g). The corresponding local conduction band edge profile, derived from TB models parametrized for multilayer h-BN, is presented in Fig.~\ref{efig_Helical_QHO} (h) exhibiting pronounced minima for electrons. The hole-counterpart from the valence band also manifests a well-defined maxima profile (not shown).

\begin{table}[b]
\caption{{The parameters of 2D QHO potentials for the helical stacking ($\theta_{12}=\theta_{23}=\theta=0.082^\circ$) and twisted monolayer-bilayer TTBN ($\theta=0.082^\circ$).}
The parameters and their unit follow the convention in Table~\ref{etable_qho_parameter}.
H-CB (H-VB) indicates QHO potential parameters for potential wells near the conduction (valence) band edge of helical stacking while M-CB (M-VB) is that of twisted monolayer-bilayer stacking.
}\label{etable_qho_parameter2}
\centering
\begin{ruledtabular}
\begin{tabular}{lccccc}
  & $m_x$  & $m_y$  & $\hbar\omega_x$  & $\hbar\omega_y$  & $E_0$  \\
\hline
H-CB & 0.496    & 0.496    & 55 & 55 & $-215$ \\  
H-VB & $-0.662$ & $-0.662$ & 32 & 32 & 148 \\  
M-CB & 0.442    & 0.525    & 50 & 82 & $-330$ \\  
M-VB & $-0.586$ & $-0.693$ & 26 & 60 & 164 \\  
\end{tabular}
\end{ruledtabular}
\end{table}

\begin{table}[b]
\caption{{Comparisons of QD state energies using TB approximation and 2D QHO models for the helical stacking and twisted monolayer-bilayer TTBNs.} Relative values to the conduction band edge $E-E_{\text{CB}}$ and to the valence band edge $E-E_{\text{VB}}$ in meV. H-CB(VB) and M-CB(VB) follow the convention in Table~\ref{etable_qho_parameter2}.}\label{etable_qho_energy2}
\centering
\begin{ruledtabular}
\begin{tabular}{lcccccccc}
$(n_x,n_y)$  & $(0,0)$ & $(1,0)$ & $(0,1)$  & $(2,0)$ & $(1,1)$  & $(0,2)$ \\
\hline
H-CB TB     & $-164$ & $-112$ & $-112$ & $-63$ & $-63$ & $-63$ \\
H-CB QHO    & $-161$ & $-107$ & $-107$ & $-52$ & $-52$ & $-52$ \\
\hline
H-VB TB     & 117 & 88 & 88 & - & - & - \\
H-VB QHO    & 115 & 83 & 83 & - & - & - \\
\hline
M-CB TB     & $-271$ & $-220$ & $-192$ & $-174$ & $-151$ & $-126$ \\
M-CB QHO    & $-264$ & $-215$ & $-182$ & $-165$ & $-133$ & $-100$ \\
\hline
M-VB TB     & 124 & 102 & 73 & - & - & - \\
M-VB QHO    & 121 & 96 & 62 & - & - & - \\
\end{tabular}
\end{ruledtabular}
\end{table}

\begin{figure}[t]
\centering
\includegraphics[width=1.0\columnwidth]{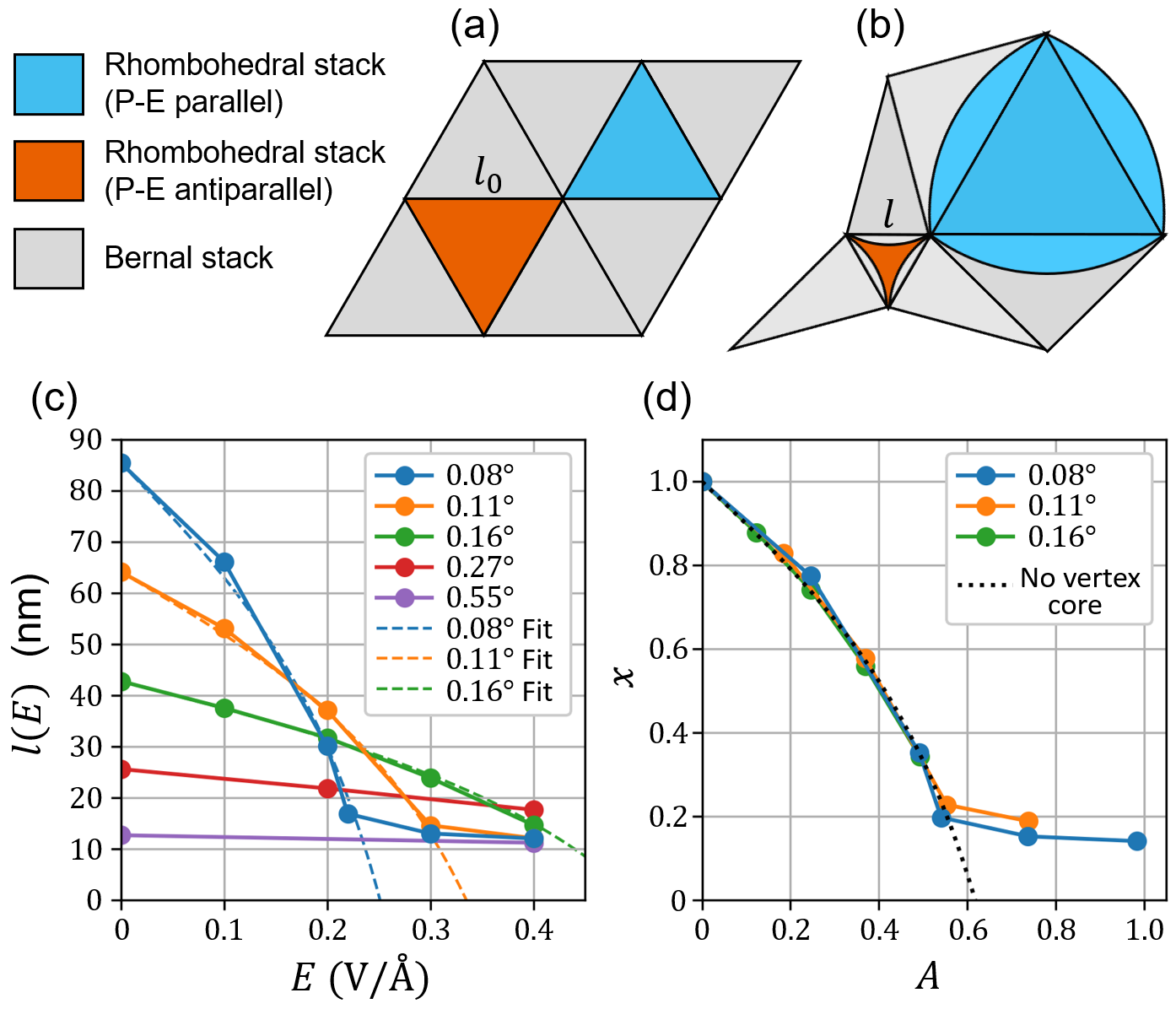}
\caption{
{Universal scaling rule for interdot distance in alternating stack.}
Domain pattern of alternating stacking TTBN of $(\theta_{12}, \theta_{23})=(-\theta, 2\theta)$ (a) without and (b) with the out-of-plane electric field. Blue (red) denotes the rhombohedral stack region whose polarization is (anti)parallel to the field, and gray denotes the Bernal stack. (c) Fully relaxed atomistic simulations on electric-field dependence of the interdot distances $l(E)$ of the Dot 2 states of $n_x = n_y = 0$ in the valence band for different twist angle $\theta$. The filled circles represent the calculated data, and dashed lines are the corresponding fitting curves. (d) The same data as in (c), rescaled using the parameters in the analytic formula, $A=aE/\theta$ and $x=l(E)/l_0$, with fitting parameter $a=0.203$ degrees$\cdot$\AA/V. ``no vertex core'' curve represents $x_{\text{min}}^{\text{no-core}} =1-A/(1-A^2)$.
}
\label{fig_QD_scale}
\end{figure}

The electronic structures and charge density distributions of the confined states are obtained from direct TB calculations for multimillion atoms.  
For the helical case, calculated LDOS projected at each QD site shown in Fig.~\ref{efig_Helical_QHO} (a) exhibits well-isolated peaks representing QD states. The first three lowest-energy conduction band states at site 1 are shown in Fig.~\ref{efig_Helical_QHO} (i). These states exhibit circular symmetry and nodal structures consistent with the isotropic 2D QHO states. Parameters characterizing the QHO potentials and the corresponding QD energy levels are provided in Tables~\ref{etable_qho_parameter2} and~\ref{etable_qho_energy2}.

Additional well-defined QD states appear at even lower-symmetric vertices, which are labeled as 2 and $2'$ in Fig.~\ref{efig_Helical_QHO} (c) and indicated by corresponding LDOS peaks in Fig.~\ref{efig_Helical_QHO} (a). These QDs form a honeycomb array [Fig.~\ref{efig_Helical_QHO} (e)], with threefold rotational equivalence around the $1'$ site. 
Under an out-of-plane field of $\left|E\right|=0.1$ V/\AA, the QD states at moved 2 and $2'$ sites [Fig.~\ref{efig_Helical_QHO} (d)] remain localized but move closer by $\sim$15 nm [Fig.~\ref{efig_Helical_QHO} (f)], demonstrating their electric reconfigurability. It is noteworthy that the LDOS of sites 1 and $1'$, which were degenerate at zero field, also split under the applied field, further verifying field-tunable coupling between QDs.

For the twisted monolayer–bilayer TTBN, as shown in Fig.~\ref{efig_MonoBi_QHO} (a), we also analyze the electronic structure, in which the LDOS peaks imply the well localized QD states. The QD sites are indicated in Fig.~\ref{efig_MonoBi_QHO} (c) depicting the fully relaxed interlayer spacing in the same spatial region of Fig.~\ref{fig_domain} (f). A zoomed-in view of the site 1 in Fig.~\ref{efig_MonoBi_QHO} (g) also reveals a distinct local maximum in interlayer spacing, again suggesting strong local confinement potential.

The conduction band energy landscape at this site [Figs.~\ref{efig_MonoBi_QHO} (h)] also shows similar features to the helical case, with deeper potential wells for electrons and holes (not shown) indicated by darker regions in the respective maps. The associated localized electronic states include the six lowest-energy states in the conduction band as shown in Fig.~\ref{efig_MonoBi_QHO} (i). The wavefunction profiles of these states closely follow the expected nodal patterns of anisotropic 2D QHO eigenfunctions. The full set of QHO potential parameters and the energy spectrum of the QD states are also listed in Tables~\ref{etable_qho_parameter2} and~\ref{etable_qho_energy2} for comparison.

In addition, the lower-symmetric vertex sites 2 and 3 labeled in Fig.~\ref{efig_MonoBi_QHO} (c) also host spatially well-isolated QD states, as indicated by the LDOS peaks in Fig.~\ref{efig_MonoBi_QHO} (a). Their density distributions are shown in Fig.~\ref{efig_MonoBi_QHO} (e). Interestingly, the energy levels of Dot 2 and Dot 3 states fall within the energy range of the 1D boundary states that interconnect the Dot 1 sites [Figs.~\ref{efig_Boundary} (c) and (f)], even though these states are spatially well separated from the boundary states. 
Under an electric field $\left|E\right|=0.03$ V/\AA, the three QD sites move closer to each other by $\sim$20 nm as shown in Fig.~\ref{efig_MonoBi_QHO} (d).
Furthermore, Dot 2 and Dot 3 states approach the 1D boundary “quantum-wire” states as shown in Fig.~\ref{efig_MonoBi_QHO} (f), where the QD and 1D states appear together in the Gaussian-weighted density distributions. This represents a distinctive aspect of electric reconfiguration in the kagome-like pattern. As described in Appendix~\ref{appendix:slideFE}, applying a stronger electric field than 0.03 V/\AA\ induces a transition to a triangular pattern, rather than further reducing the interdot distance. For a similar reason, the QD motion in the kagome-like pattern is prominently more susceptible to the applied electric field.

\section{A Scaling Rule of Inder-dot Distances in Alternated stacking Systems}\label{sec:appendix_QD_scale}

In this appendix, we derive a single scaling relation between the electric-field and twist-angle dependence of interdot distance in the alternated stack TTBN, which provides practical information to estimate the critical electric field $E_C$ to achieve the interdot coupling for different moir\'e periods. 
Fig.~\ref{fig_QD_scale} (c) shows the electric-field dependence of the interdot distances $l$ of the highest Dot 2 states of $n_x = n_y = 0$ in the valence band of fully relaxed atomic structures of alternated stack $(\theta_{12}, \theta_{23})=(-\theta, 2\theta)$ with different $\theta$.
It shows that the critical electric field of $E_C$ required to saturate interdot spacing to $l_C$ increases as the twist angle increases. The saturation can be well understood once we consider the hardcore effects on the vertex region where the elastic compression energy increases sharply as interdot distance decreases. Due to the local hardcore effects at the vertex, the $l_C$ does not depend on the twist angle when the twist angle is small. 

To elucidate these numerical simulation results phenomenologically, we formulated the total energy change ($\Delta\varepsilon$) under an applied field ($E$) in terms of the normalized interdot spacing $x(\theta,E)=l(\theta,E)/l_0$ where $l_0\equiv l(\theta,E=0)=1.23/\theta$ is an angle-dependent interdot spacing in the absence of $E$ (its unit is~\AA/rad). $l_0$ without $E$ and $l$ with $E$ are denoted in Figs.~\ref{fig_QD_scale}(a) and (b), repsectively. 
By using the domain geometries without and with $E$ [Figs.~\ref{fig_QD_scale} (a) and (b)],  $\Delta\varepsilon$ upon the applied $E$ can be written as,
\begin{eqnarray}
\Delta\varepsilon &=& 12\varepsilon_{W}l_0\sqrt{1+\frac{1}{3}(x-1)^2} 
- 4\sqrt{3}\delta_{S}l_0^2 (1-x) \nonumber \\
&&- \frac{6\delta_{S}^2}{\varepsilon_W}l_0^3(1-x)^2 
+ f_c(x),
\label{eq_scale1}
\end{eqnarray}
where $\varepsilon_W$ denotes the domain-wall energy per unit length, and $\delta_S$ the stacking energy difference per unit area between Bernal (nonpolar) and rhombohedral (polar) domains. We note $\delta_S\sim\pm P E$ for the red and blue regions in Fig.~\ref{fig_QD_scale} where $P$ is the absolute magnitude of the layer polarization of the rhombohedral domain. 
The first term represents the elastic energy cost arising from the increased total length of domain walls with increasing field and the second term describes the energy gain from the field-induced expansion or shrinkage of the rhombohedral domains. The third term proportional to $l_0^3$ accounts for the warping of domain walls, and the last term of $f_c(x)$ introduces a hardcore vertex energy that diverges when $l$ approaches $l_C$. Here, we simply assume $f_c(x)=\varepsilon_{c}/(x-x_C)$ where $x_C=l_C/l_0$ and $\varepsilon_c$ is a constant.       

Equation~\eqref{eq_scale1} can be recast in dimensionless form as 
\begin{eqnarray}
\frac{\Delta\varepsilon}{2\varepsilon_{W}l_0} 
&\approx& \left(1-\frac{a^2E^2}{\theta^2}\right)(1-x)^2 - 2\frac{aE}{\theta}(1-x) \nonumber \\
& &+ \frac{b\theta}{x-x_C} ,   
\label{eq_scale2}
\end{eqnarray}
where $a$ and $b$ are independent of twist angle and field strength. The equilibrium interdot spacing is obtained by minimizing Eq.~\eqref{eq_scale2} with respect to $x$. 

Comparison against the simulated data $l(\theta,E)$ is performed without the hardcore vertex effect term $b\theta/(x-x_C)$ for further simplicity. Then, Eq.~(\ref{eq_scale2}) becomes a much simplified form of $(1-A^2)(1-x)^2 -2A(1-x)$, where $A\equiv aE/\theta$, thus directly giving a minimization condition $x_{\text{min}}^{\text{no-core}}(A)=1-A/(1-A^2)$.
We fitted the parameter $a$ against the simulated $l(\theta,E)$ only for angles smaller than 0.2 degrees and for $E$ smaller than the saturation regime.  
The single parameter $a=0.203$ degrees$\cdot$\AA/V\ reproduces 
all the numerical data with different twist angles accurately, as shown by the dashed lines in Fig.~\ref{fig_QD_scale} (c). Furthermore, when the interdot distances and electric fields are rescaled in terms of the variables $x$ and $A$, all data before the saturation collapse onto a single scaling curve by $x_{\text{min}}^{\text{no-core}}$ [Fig.~\ref{fig_QD_scale} (d)].
These results establish a single, well-defined scaling rule for the variation of domain area with electric field that unifies all calculated results across varying twist angles. With this, for all considered angles, the critical $A_C$ for saturating $x_C$ is almost constant for all angles so that $E_C=A_C \theta/a\sim \theta$, explaining our calculations very well.

\section{Inhomogeneity effect}\label{sec:appendix_inhomo}
Inhomogeneity of the sample is often thought to hinder the reproducibility of phenomena in moir\'e systems. The inhomogeneity may come from two different materials parameters, namely twist angle disorder and local variations of strains. However, as demonstrated in the previous study on TTG~\cite{Park2025Nature}, the formation of stacking domain patterns is robust over a finite range of twist-angle disorder and incommensuration. Even under these conditions, the trilayer lattice tends to self-organize into locally commensurate domain structures~\cite{Park2025Nature}, which would generate almost identical QD states at each vertex. 
In particular, when we examine the sequence of twist angles used in Fig.~\ref{fig_QD_scale} (c), the QD states at $E=0$ in the alternate stack persist and the corresponding energy levels and spacing vary by only a few meV, except for shallowly bound states.
Moreover, to check the effect of strain on the QD energy levels, we consider a 0.5 \% uniaxial tensile strain along either the armchair or zigzag direction for the fully relaxed $\theta=0.08^\circ$ alternated stack structure without performing additional relaxation. All QD states remain, with a uniform shift in their absolute energies, while the level spacings stay nearly unchanged. The overall band gap decreases by about 40 meV, and the threefold degeneracy of the equivalent Dot 2 sites is lifted by only a few meV. 
Together, these results confirm that the QD energy levels and couplings are robust against small twist-angle disorder and weak strain, ensuring the experimental feasibility and reproducibility of the proposed field-controlled QD arrays. 
Furthermore, experimental advances in reducing moir\'e sample inhomogeneity~\cite{kapfer2023science,wu2023interface} will strengthen the feasibility of implementing our proposal.

\bibliography{TTBN_PRX}

@misc{inhouse,
     howpublished = {\url{https://github.com/KIAS-CMT/DFT-U-V}}
}

@article{Naik2018PRL,
  title = {Ultraflatbands and Shear Solitons in Moir\'e Patterns of Twisted Bilayer Transition Metal Dichalcogenides},
  author = {Naik, Mit H. and Jain, Manish},
  journal = {Phys. Rev. Lett.},
  volume = {121},
  issue = {26},
  pages = {266401},
  numpages = {6},
  year = {2018},
  month = {Dec},
  publisher = {American Physical Society},
  doi = {10.1103/PhysRevLett.121.266401},
  url = {https://link.aps.org/doi/10.1103/PhysRevLett.121.266401}
}

@article{Koshino2018PRX,
  title = {Maximally Localized Wannier Orbitals and the Extended Hubbard Model for Twisted Bilayer Graphene},
  author = {Koshino, Mikito and Yuan, Noah F. Q. and Koretsune, Takashi and Ochi, Masayuki and Kuroki, Kazuhiko and Fu, Liang},
  journal = {Phys. Rev. X},
  volume = {8},
  issue = {3},
  pages = {031087},
  numpages = {12},
  year = {2018},
  month = {Sep},
  publisher = {American Physical Society},
  doi = {10.1103/PhysRevX.8.031087},
  url = {https://link.aps.org/doi/10.1103/PhysRevX.8.031087}
}

@article{Weston2022NatureNano,
	Author = {Weston, Astrid and Castanon, Eli G. and Enaldiev, Vladimir and Ferreira, F{\'a}bio and Bhattacharjee, Shubhadeep and Xu, Shuigang and Corte-Le{\'o}n, H{\'e}ctor and Wu, Zefei and Clark, Nicholas and Summerfield, Alex and Hashimoto, Teruo and Gao, Yunze and Wang, Wendong and Hamer, Matthew and Read, Harriet and Fumagalli, Laura and Kretinin, Andrey V. and Haigh, Sarah J. and Kazakova, Olga and Geim, A. K. and Fal'ko, Vladimir I. and Gorbachev, Roman},
	Doi = {10.1038/s41565-022-01072-w},
	Journal = {Nature Nanotechnol.},
	Number = {4},
	Pages = {390--395},
	Title = {Interfacial ferroelectricity in marginally twisted 2D semiconductors},
	Url = {https://doi.org/10.1038/s41565-022-01072-w},
	Volume = {17},
	Year = {2022}
}

@article{Yang2021PRB,
  title = {{Ab initio study of lattice dynamics of group {IV} semiconductors using pseudohybrid functionals for extended Hubbard interactions}},
  author = {Yang, Wooil and Jhi, Seung-Hoon and Lee, Sang-Hoon and Son, Young-Woo},
  journal = {Phys. Rev. B},
  volume = {104},
  issue = {10},
  pages = {104313},
  numpages = {12},
  year = {2021},
  month = {Sep},
  publisher = {American Physical Society},
  doi = {10.1103/PhysRevB.104.104313},
  url = {https://link.aps.org/doi/10.1103/PhysRevB.104.104313}
}

@article{Rubio2020PRB,
  title = {Parameter-free hybridlike functional based on an extended \text{Hubbard} model: $\mathrm{DFT}+\text{U}+\text{V}$},
  author = {Tancogne-Dejean, Nicolas and Rubio, Angel},
  journal = {Phys. Rev. B},
  volume = {102},
  issue = {15},
  pages = {155117},
  numpages = {11},
  year = {2020},
  month = {Oct},
  publisher = {American Physical Society},
  doi = {10.1103/PhysRevB.102.155117},
  url = {https://link.aps.org/doi/10.1103/PhysRevB.102.155117}
}

@article{Timrov2020prb,
  title = {Pulay forces in density-functional theory with extended Hubbard functionals: From nonorthogonalized to orthogonalized manifolds},
  author = {Timrov, Iurii and Aquilante, Francesco and Binci, Luca and Cococcioni, Matteo and Marzari, Nicola},
  journal = {Phys. Rev. B},
  volume = {102},
  issue = {23},
  pages = {235159},
  numpages = {15},
  year = {2020},
  month = {Dec},
  publisher = {American Physical Society},
  doi = {10.1103/PhysRevB.102.235159},
  url = {https://link.aps.org/doi/10.1103/PhysRevB.102.235159}
}

@article{Park2025Nature,
	title = {Unconventional domain tessellations in moir\'{e}-of-moir\'{e} lattices},
	volume = {641},
	issn = {0028-0836, 1476-4687},
	url = {https://www.nature.com/articles/s41586-025-08932-0},
	doi = {10.1038/s41586-025-08932-0},
	pages = {896--903},
	number = {8064},
	journaltitle = {Nature},
	journal = {Nature},
	shortjournal = {Nature},
	author = {Park, Daesung and Park, Changwon and Yananose, Kunihiro and Ko, Eunjung and Kim, Byunghyun and Engelke, Rebecca and Zhang, Xi and Davydov, Konstantin and Green, Matthew and Kim, Hyun-Mi and Park, Sang Hwa and Lee, Jae Heon and Kim, Seul-Gi and Kim, Hyeongkeun and Watanabe, Kenji and Taniguchi, Takashi and Yang, Sang Mo and Wang, Ke and Kim, Philip and Son, Young-Woo and Yoo, Hyobin},
	urldate = {2025-05-27},
	date = {2025-05-22},
	  year = {2025},
	langid = {english},
}

@article{Xian2019NL,
author = {Xian, Lede and Kennes, Dante M. and Tancogne-Dejean, Nicolas and Altarelli, Massimo and Rubio, Angel},
title = {Multiflat Bands and Strong Correlations in Twisted Bilayer Boron Nitride: Doping-Induced Correlated Insulator and Superconductor},
journal = {Nano Lett.},
volume = {19},
number = {8},
pages = {4934-4940},
year = {2019}
}

@article{Naik2020PRB,
  title = {Origin and evolution of ultraflat bands in twisted bilayer transition metal dichalcogenides: Realization of triangular quantum dots},
  author = {Naik, Mit H. and Kundu, Sudipta and Maity, Indrajit and Jain, Manish},
  journal = {Phys. Rev. B},
  volume = {102},
  issue = {7},
  pages = {075413},
  numpages = {11},
  year = {2020},
  month = {Aug}
}

@article{Li2024PRB,
  title = {Moir\'e flat bands and antiferroelectric domains in lattice relaxed twisted bilayer hexagonal boron nitride under perpendicular electric fields},
  author = {Li, Fengping and Lee, Dongkyu and Leconte, Nicolas and Javvaji, Srivani and Kim, Young Duck and Jung, Jeil},
  journal = {Phys. Rev. B},
  volume = {110},
  issue = {15},
  pages = {155419},
  numpages = {13},
  year = {2024},
  month = {Oct}
}

@Article{Kennes2021NP,
author={Kennes, Dante M.
and Claassen, Martin
and Xian, Lede
and Georges, Antoine
and Millis, Andrew J.
and Hone, James
and Dean, Cory R.
and Basov, D. N.
and Pasupathy, Abhay N.
and Rubio, Angel},
title={Moir{\'e} heterostructures as a condensed-matter quantum simulator},
journal={Nat. Phys.},
year={2021},
month={Feb},
day={01},
volume={17},
number={2},
pages={155-163}
}

@Article{Enaldiev2022npj2D,
author={Enaldiev, V. V.
and Ferreira, F.
and McHugh, J. G.
and Fal'ko, Vladimir I.},
title={Self-organized quantum dots in marginally twisted {MoSe$_2$/WSe$_2$} and {MoS$_2$/WS$_2$} bilayers},
journal={npj 2D Mater. Appl.},
year={2022},
month={Oct},
day={23},
volume={6},
number={1},
pages={74}
}

@Article{Montblanch2023NatNano,
author={Montblanch, Alejandro R.-P.
and Barbone, Matteo
and Aharonovich, Igor
and Atat{\"u}re, Mete
and Ferrari, Andrea C.},
title={Layered materials as a platform for quantum technologies},
journal={Nat. Nanotechnol.},
year={2023},
month={Jun},
day={01},
volume={18},
number={6},
pages={555-571},
issn={1748-3395}
}

@Article{Wang2018NatNano,
author={Wang, Ke
and De Greve, Kristiaan
and Jauregui, Luis A.
and Sushko, Andrey
and High, Alexander
and Zhou, You
and Scuri, Giovanni
and Taniguchi, Takashi
and Watanabe, Kenji
and Lukin, Mikhail D.
and Park, Hongkun
and Kim, Philip},
title={Electrical control of charged carriers and excitons in atomically thin materials},
journal={Nat. Nanotechnol.},
year={2018},
month={Feb},
day={01},
volume={13},
number={2},
pages={128-132},
issn={1748-3395}
}

@misc{nakatsuji2025arxiv,
	title = {Moir\'{e} Band Engineering in Twisted Trilayer {WSe$_2$}},
	url = {http://arxiv.org/abs/2504.20449},
	doi = {10.48550/arXiv.2504.20449},
	number = {{arXiv}:2504.20449},
	publisher = {{arXiv}},
	author = {Nakatsuji, Naoto and Kawakami, Takuto and Tateishi, Hayato and Kato, Koichiro and Koshino, Mikito},
	urldate = {2025-05-27},
	date = {2025-04-29},
	eprinttype = {arxiv},
	eprint = {2504.20449},
        year={2025},
}

@article{zhao2020PRL,
	title = {Formation of Bloch Flat Bands in Polar Twisted Bilayers without Magic Angles},
	volume = {124},
	issn = {0031-9007, 1079-7114},
	url = {https://link.aps.org/doi/10.1103/PhysRevLett.124.086401},
	doi = {10.1103/PhysRevLett.124.086401},
	pages = {086401},
	number = {8},
	journaltitle = {Physical Review Letters},
	shortjournal = {Phys. Rev. Lett.},
	journal = {Phys. Rev. Lett.},
    author = {Zhao, Xing-Ju and Yang, Yang and Zhang, Dong-Bo and Wei, Su-Huai},
	urldate = {2025-12-13},
	date = {2020-02-26},
	year = {2020},
	langid = {english},
}

@article{Grose2020APLPh,
    author = {Gro{\ss}e, Jan and von Helversen, Martin and Koulas-Simos, Aris and Hermann, Martin and Reitzenstein, Stephan},
    title = {Development of site-controlled quantum dot arrays acting as scalable sources of indistinguishable photons},
    journal = {APL Photonics},
    volume = {5},
    number = {9},
    pages = {096107},
    year = {2020},
    month = {09}
}

@article{Chung2013PRB,
  title = {Charge stability of a triple quantum dot with a finite tunnel coupling},
  author = {Lee, Soo-Young and Chung, Yunchul},
  journal = {Phys. Rev. B},
  volume = {87},
  issue = {4},
  pages = {045302},
  numpages = {7},
  year = {2013},
  month = {Jan},
  publisher = {American Physical Society}
}

@article{Chung2013PRL,
  title = {Charge Frustration in a Triangular Triple Quantum Dot},
  author = {Seo, M. and Choi, H. K. and Lee, S.-Y. and Kim, N. and Chung, Y. and Sim, H.-S. and Umansky, V. and Mahalu, D.},
  journal = {Phys. Rev. Lett.},
  volume = {110},
  issue = {4},
  pages = {046803},
  numpages = {5},
  year = {2013},
  month = {Jan},
  publisher = {American Physical Society}
}

@Article{Hamo2016Nature,
author={Hamo, A.
and Benyamini, A.
and Shapir, I.
and Khivrich, I.
and Waissman, J.
and Kaasbjerg, K.
and Oreg, Y.
and von Oppen, F.
and Ilani, S.},
title={Electron attraction mediated by Coulomb repulsion},
journal={Nature},
year={2016},
month={Jul},
day={01},
volume={535},
number={7612},
pages={395-400},
issn={1476-4687}
}

@Article{Dehollain2020Nature,
author={Dehollain, J. P.
and Mukhopadhyay, U.
and Michal, V. P.
and Wang, Y.
and Wunsch, B.
and Reichl, C.
and Wegscheider, W.
and Rudner, M. S.
and Demler, E.
and Vandersypen, L. M. K.},
title={Nagaoka ferromagnetism observed in a quantum dot plaquette},
journal={Nature},
year={2020},
month={Mar},
day={01},
volume={579},
number={7800},
pages={528-533},
issn={1476-4687}
}

@article{Loss1998PRA,
  title = {Quantum computation with quantum dots},
  author = {Loss, Daniel and DiVincenzo, David P.},
  journal = {Phys. Rev. A},
  volume = {57},
  issue = {1},
  pages = {120--126},
  numpages = {0},
  year = {1998},
  month = {Jan},
  publisher = {American Physical Society}
}

@article{Hsieh2012RPP,
year = {2012},
month = {oct},
publisher = {IOP Publishing},
volume = {75},
number = {11},
pages = {114501},
author = {Hsieh, Chang-Yu and Shim, Yun-Pil and Korkusinski, Marek and Hawrylak, Pawel},
title = {Physics of lateral triple quantum-dot molecules with controlled electron numbers},
journal = {Rep. Prog. Phys.}
}

@Article{Philips2022Nature,
author={Philips, Stephan G. J.
and M{\k{a}}dzik, Mateusz T.
and Amitonov, Sergey V.
and de Snoo, Sander L.
and Russ, Maximilian
and Kalhor, Nima
and Volk, Christian
and Lawrie, William I. L.
and Brousse, Delphine
and Tryputen, Larysa
and Wuetz, Brian Paquelet
and Sammak, Amir
and Veldhorst, Menno
and Scappucci, Giordano
and Vandersypen, Lieven M. K.},
title={Universal control of a six-qubit quantum processor in silicon},
journal={Nature},
year={2022},
month={Sep},
day={01},
volume={609},
number={7929},
pages={919-924},
issn={1476-4687}
}

@Article{Borsoi2024NatNano,
author={Borsoi, Francesco
and Hendrickx, Nico W.
and John, Valentin
and Meyer, Marcel
and Motz, Sayr
and van Riggelen, Floor
and Sammak, Amir
and de Snoo, Sander L.
and Scappucci, Giordano
and Veldhorst, Menno},
title={Shared control of a 16{\thinspace}semiconductor quantum dot crossbar array},
journal={Nat. Nanotechnol.},
year={2024},
month={Jan},
day={01},
volume={19},
number={1},
pages={21-27}
}

@article{Burkard2023RMP,
  title = {Semiconductor spin qubits},
  author = {Burkard, Guido and Ladd, Thaddeus D. and Pan, Andrew and Nichol, John M. and Petta, Jason R.},
  journal = {Rev. Mod. Phys.},
  volume = {95},
  issue = {2},
  pages = {025003},
  numpages = {58},
  year = {2023},
  month = {Jun},
  publisher = {American Physical Society}
}

@article{Yan2019AdvPhys,
author = {Linghao Yan and Peter Liljeroth},
title = {Engineered electronic states in atomically precise artificial lattices and graphene nanoribbons},
journal = {Adv. Phys.: X},
volume = {4},
number = {1},
pages = {1651672},
year = {2019},
publisher = {Taylor \& Francis}
}

@article{Pan2015PRL,
  title = {Reconfigurable Quantum-Dot Molecules Created by Atom Manipulation},
  author = {Pan, Yi and Yang, Jianshu and Erwin, Steven C. and Kanisawa, Kiyoshi and F\"olsch, Stefan},
  journal = {Phys. Rev. Lett.},
  volume = {115},
  issue = {7},
  pages = {076803},
  numpages = {5},
  year = {2015},
  month = {Aug},
  publisher = {American Physical Society}
}

@article{javvaji_ab_2025,
	title = {\textit{Ab initio} tight-binding models for mono- and bilayer hexagonal boron nitride (h-{BN})},
	volume = {9},
	issn = {2475-9953},
	url = {https://link.aps.org/doi/10.1103/PhysRevMaterials.9.024004},
	doi = {10.1103/PhysRevMaterials.9.024004},
	pages = {024004},
	number = {2},
	journaltitle = {Physical Review Materials},
	shortjournal = {Phys. Rev. Materials},
	journal = {Phys. Rev. Materials},
        author = {Javvaji, Srivani and Li, Fengping and Jung, Jeil},
	urldate = {2025-07-26},
	date = {2025-02-25},
	year = {2025},
	langid = {english},
}

@article{Constantinescu2013PRL,
  title = {Stacking in Bulk and Bilayer Hexagonal Boron Nitride},
  author = {Constantinescu, Gabriel and Kuc, Agnieszka and Heine, Thomas},
  journal = {Phys. Rev. Lett.},
  volume = {111},
  issue = {3},
  pages = {036104},
  numpages = {5},
  year = {2013},
  month = {Jul},
  publisher = {American Physical Society}
}

@article{Warner2010ACSNano,
author = {Warner, Jamie H. and R{\"u}mmeli, Mark H. and Bachmatiuk, Alicja and B{\"u}chner, Bernd},
title = {Atomic Resolution Imaging and Topography of Boron Nitride Sheets Produced by Chemical Exfoliation},
journal = {ACS Nano},
volume = {4},
number = {3},
pages = {1299-1304},
year = {2010}
}

@article{Vizner2021Science,
	title = {Interfacial ferroelectricity by van der {Waals} sliding},
	volume = {372},
	number = {6549},
	urldate = {2025-02-06},
	journal = {Science},
	author = {Vizner Stern, M. and Waschitz, Y. and Cao, W. and Nevo, I. and Watanabe, K. and Taniguchi, T. and Sela, E. and Urbakh, M. and Hod, O. and Ben Shalom, M.},
	month = jun,
	year = {2021},
	pages = {1462--1466},
}

@article{Yasuda2021Science,
	title = {Stacking-engineered ferroelectricity in bilayer boron nitride},
	volume = {372},
	number = {6549},
	urldate = {2025-02-06},
	journal = {Science},
	author = {Yasuda, Kenji and Wang, Xirui and Watanabe, Kenji and Taniguchi, Takashi and Jarillo-Herrero, Pablo},
	month = jun,
	year = {2021},
	pages = {1458--1462},
}

@Article{Caldwell2019NatRevMat,
author={Caldwell, Joshua D.
and Aharonovich, Igor
and Cassabois, Guillaume
and Edgar, James H.
and Gil, Bernard
and Basov, D. N.},
title={Photonics with hexagonal boron nitride},
journal={Nat. Rev. Mater.},
year={2019},
month={Aug},
day={01},
volume={4},
number={8},
pages={552-567}
}

@Article{Su2022NatMat,
author={Su, Cong
and Zhang, Fang
and Kahn, Salman
and Shevitski, Brian
and Jiang, Jingwei
and Dai, Chunhui
and Ungar, Alex
and Park, Ji-Hoon
and Watanabe, Kenji
and Taniguchi, Takashi
and Kong, Jing
and Tang, Zikang
and Zhang, Wenqing
and Wang, Feng
and Crommie, Michael
and Louie, Steven G.
and Aloni, Shaul
and Zettl, Alex},
title={Tuning colour centres at a twisted hexagonal boron nitride interface},
journal={Nat. Mater.},
year={2022},
month={Aug},
day={01},
volume={21},
number={8},
pages={896-902}
}

@Article{Tran2016NatNano,
author={Tran, Toan Trong
and Bray, Kerem
and Ford, Michael J.
and Toth, Milos
and Aharonovich, Igor},
title={Quantum emission from hexagonal boron nitride monolayers},
journal={Nat. Nanotechnol.},
year={2016},
month={Jan},
day={01},
volume={11},
number={1},
pages={37-41}
}

@article{Liu2022MQT,
year = {2022},
month = {jul},
publisher = {IOP Publishing},
volume = {2},
number = {3},
pages = {032002},
author = {Liu, Wei and Guo, Nai-Jie and Yu, Shang and Meng, Yu and Li, Zhi-Peng and Yang, Yuan-Ze and Wang, Zhao-An and Zeng, Xiao-Dong and Xie, Lin-Ke and Li, Qiang and Wang, Jun-Feng and Xu, Jin-Shi and Wang, Yi-Tao and Tang, Jian-Shun and Li, Chuan-Feng and Guo, Guang-Can},
title = {Spin-active defects in hexagonal boron nitride},
journal = {Mater. Quantum Technol.}
}

@Article{Gong2024NatComm,
author={Gong, Ruotian
and Du, Xinyi
and Janzen, Eli
and Liu, Vincent
and Liu, Zhongyuan
and He, Guanghui
and Ye, Bingtian
and Li, Tongcang
and Yao, Norman Y.
and Edgar, James H.
and Henriksen, Erik A.
and Zu, Chong},
title={Isotope engineering for spin defects in van der Waals materials},
journal={Nat. Commun.},
year={2024},
month={Jan},
day={02},
volume={15},
number={1},
pages={104}
}

@Article{Rizzato2023NatComm,
author={Rizzato, Roberto
and Schalk, Martin
and Mohr, Stephan
and Hermann, Jens C.
and Leibold, Joachim P.
and Bruckmaier, Fleming
and Salvitti, Giovanna
and Qian, Chenjiang
and Ji, Peirui
and Astakhov, Georgy V.
and Kentsch, Ulrich
and Helm, Manfred
and Stier, Andreas V.
and Finley, Jonathan J.
and Bucher, Dominik B.},
title={Extending the coherence of spin defects in {hBN} enables advanced qubit control and quantum sensing},
journal={Nat. Commun.},
year={2023},
month={Aug},
day={22},
volume={14},
number={1},
pages={5089}
}

@article{Okada2011PRB,
  title = {Gate-controlled carrier injection into hexagonal boron nitride},
  author = {Otani, Minoru and Okada, Susumu},
  journal = {Phys. Rev. B},
  volume = {83},
  issue = {7},
  pages = {073405},
  numpages = {4},
  year = {2011},
  month = {Feb},
  publisher = {American Physical Society}
}

@Article{Ladd2010Nature,
author={Ladd, T. D.
and Jelezko, F.
and Laflamme, R.
and Nakamura, Y.
and Monroe, C.
and O'Brien, J. L.},
title={Quantum computers},
journal={Nature},
year={2010},
month={Mar},
day={01},
volume={464},
number={7285},
pages={45-53}
}

@article{Bruzewicz2019APR,
    author = {Bruzewicz, Colin D. and Chiaverini, John and McConnell, Robert and Sage, Jeremy M.},
    title = {Trapped-ion quantum computing: Progress and challenges},
    journal = {Appl. Phys. Rev.},
    volume = {6},
    number = {2},
    pages = {021314},
    year = {2019},
    month = {05}
}

@article{Monroe2021Nature,
  title = {Programmable quantum simulations of spin systems with trapped ions},
  author = {Monroe, C. and Campbell, W. C. and Duan, L.-M. and Gong, Z.-X. and Gorshkov, A. V. and Hess, P. W. and Islam, R. and Kim, K. and Linke, N. M. and Pagano, G. and Richerme, P. and Senko, C. and Yao, N. Y.},
  journal = {Rev. Mod. Phys.},
  volume = {93},
  issue = {2},
  pages = {025001},
  numpages = {57},
  year = {2021},
  month = {Apr}
}

@Article{Bluvstein2022Nature,
author={Bluvstein, Dolev
and Levine, Harry
and Semeghini, Giulia
and Wang, Tout T.
and Ebadi, Sepehr
and Kalinowski, Marcin
and Keesling, Alexander
and Maskara, Nishad
and Pichler, Hannes
and Greiner, Markus
and Vuleti{\'{c}}, Vladan
and Lukin, Mikhail D.},
title={A quantum processor based on coherent transport of entangled atom arrays},
journal={Nature},
year={2022},
month={Apr},
day={01},
volume={604},
number={7906},
pages={451-456}
}

@Article{Ebadi2021Nature,
author={Ebadi, Sepehr
and Wang, Tout T.
and Levine, Harry
and Keesling, Alexander
and Semeghini, Giulia
and Omran, Ahmed
and Bluvstein, Dolev
and Samajdar, Rhine
and Pichler, Hannes
and Ho, Wen Wei
and Choi, Soonwon
and Sachdev, Subir
and Greiner, Markus
and Vuleti{\'{c}}, Vladan
and Lukin, Mikhail D.},
title={Quantum phases of matter on a 256-atom programmable quantum simulator},
journal={Nature},
year={2021},
month={Jul},
day={01},
volume={595},
number={7866},
pages={227-232}
}

@article{Nathalie2021Science,
author = {Nathalie P. de Leon  and Kohei M. Itoh  and Dohun Kim  and Karan K. Mehta  and Tracy E. Northup  and Hanhee Paik  and B. S. Palmer  and N. Samarth  and Sorawis Sangtawesin  and D. W. Steuerman },
title = {Materials challenges and opportunities for quantum computing hardware},
journal = {Science},
volume = {372},
number = {6539},
pages = {eabb2823},
year = {2021}
}

@Article{Andrei2021NatRevMat,
author={Andrei, Eva Y.
and Efetov, Dmitri K.
and Jarillo-Herrero, Pablo
and MacDonald, Allan H.
and Mak, Kin Fai
and Senthil, T.
and Tutuc, Emanuel
and Yazdani, Ali
and Young, Andrea F.},
title={The marvels of moir{\'e} materials},
journal={Nat. Rev. Mater.},
year={2021},
month={Mar},
day={01},
volume={6},
number={3},
pages={201-206}
}

@article{Cao_correlated_2018Nature,
	title = {Correlated insulator behaviour at half-filling in magic-angle graphene superlattices},
	volume = {556},
	number = {7699},
	urldate = {2025-02-06},
	journal = {Nature},
	author = {Cao, Yuan and Fatemi, Valla and Demir, Ahmet and Fang, Shiang and Tomarken, Spencer L. and Luo, Jason Y. and Sanchez-Yamagishi, Javier D. and Watanabe, Kenji and Taniguchi, Takashi and Kaxiras, Efthimios and Ashoori, Ray C. and Jarillo-Herrero, Pablo},
	month = apr,
	year = {2018},
	pages = {80--84},
}

@article{Bistritzer2011PNAS,
author = {Rafi Bistritzer  and Allan H. MacDonald },
title = {Moir\'{e} bands in twisted double-layer graphene},
journal = {Proc. Natl. Acad. Sci. U.S.A.},
volume = {108},
number = {30},
pages = {12233-12237},
year = {2011}
}

@article{Lopes2012PRB,
  title = {Continuum model of the twisted graphene bilayer},
  author = {Lopes dos Santos, J. M. B. and Peres, N. M. R. and Castro Neto, A. H.},
  journal = {Phys. Rev. B},
  volume = {86},
  issue = {15},
  pages = {155449},
  numpages = {12},
  year = {2012},
  month = {Oct}
}

@article{Trambly2010Nanolett,
	title = {Localization of {Dirac} {Electrons} in {Rotated} {Graphene} {Bilayers}},
	volume = {10},
	number = {3},
	urldate = {2024-01-02},
	journal = {Nano Lett.},
	author = {Trambly De Laissardi{\`e}re, G. and Mayou, D. and Magaud, L.},
	month = mar,
	year = {2010},
	pages = {804--808},
}

@Article{VanWinkle2024NatNano,
author={Van Winkle, Madeline
and Dowlatshahi, Nikita
and Khaloo, Nikta
and Iyer, Mrinalni
and Craig, Isaac M.
and Dhall, Rohan
and Taniguchi, Takashi
and Watanabe, Kenji
and Bediako, D. Kwabena},
title={Engineering interfacial polarization switching in van der Waals multilayers},
journal={Nat. Nanotechnol.},
year={2024},
month={Jun},
day={01},
volume={19},
number={6},
pages={751-757}
}

@Article{Craig2024NatMat,
author={Craig, Isaac M.
and Van Winkle, Madeline
and Groschner, Catherine
and Zhang, Kaidi
and Dowlatshahi, Nikita
and Zhu, Ziyan
and Taniguchi, Takashi
and Watanabe, Kenji
and Griffin, Sin{\'e}ad M.
and Bediako, D. Kwabena},
title={Local atomic stacking and symmetry in twisted graphene trilayers},
journal={Nat. Mater.},
year={2024},
month={Mar},
day={01},
volume={23},
number={3},
pages={323-330}
}

@article{Rebeca2018Science,
author = {Rebeca Ribeiro-Palau  and Changjian Zhang  and Kenji Watanabe  and Takashi Taniguchi  and James Hone  and Cory R. Dean },
title = {Twistable electronics with dynamically rotatable heterostructures},
journal = {Science},
volume = {361},
number = {6403},
pages = {690-693},
year = {2018}
}

@Article{Tang2024Nature,
author={Tang, Haoning
and Wang, Yiting
and Ni, Xueqi
and Watanabe, Kenji
and Taniguchi, Takashi
and Jarillo-Herrero, Pablo
and Fan, Shanhui
and Mazur, Eric
and Yacoby, Amir
and Cao, Yuan},
title={On-chip multi-degree-of-freedom control of two-dimensional materials},
journal={Nature},
year={2024},
month={Aug},
day={01},
volume={632},
number={8027},
pages={1038-1044}
}

@article{Nakatsuji2023PRX,
  title = {Multiscale Lattice Relaxation in General Twisted Trilayer Graphenes},
  author = {Nakatsuji, Naoto and Kawakami, Takuto and Koshino, Mikito},
  journal = {Phys. Rev. X},
  volume = {13},
  issue = {4},
  pages = {041007},
  numpages = {20},
  year = {2023},
  month = {Oct},
  publisher = {American Physical Society}
}

@article{QE1_2009,
	title = {{QUANTUM} {ESPRESSO}: a modular and open-source software project for quantum simulations of materials},
	volume = {21},
	shorttitle = {{QUANTUM} {ESPRESSO}},
	number = {39},
	urldate = {2025-02-07},
	journal = {J. Phys.: Condens. Matter},
	author = {Giannozzi, Paolo and Baroni, Stefano and Bonini, Nicola and Calandra, Matteo and Car, Roberto and Cavazzoni, Carlo and Ceresoli, Davide and Chiarotti, Guido L and Cococcioni, Matteo and Dabo, Ismaila and Dal Corso, Andrea and De Gironcoli, Stefano and Fabris, Stefano and Fratesi, Guido and Gebauer, Ralph and Gerstmann, Uwe and Gougoussis, Christos and Kokalj, Anton and Lazzeri, Michele and Martin-Samos, Layla and Marzari, Nicola and Mauri, Francesco and Mazzarello, Riccardo and Paolini, Stefano and Pasquarello, Alfredo and Paulatto, Lorenzo and Sbraccia, Carlo and Scandolo, Sandro and Sclauzero, Gabriele and Seitsonen, Ari P and Smogunov, Alexander and Umari, Paolo and Wentzcovitch, Renata M},
	month = sep,
	year = {2009},
	pages = {395502},
}

@article{QE2_2017,
	title = {Advanced capabilities for materials modelling with {Quantum} {ESPRESSO}},
	volume = {29},
	number = {46},
	urldate = {2025-02-07},
	journal = {J. Phys.: Condens. Matter},
	author = {Giannozzi, P and Andreussi, O and Brumme, T and Bunau, O and Buongiorno Nardelli, M and Calandra, M and Car, R and Cavazzoni, C and Ceresoli, D and Cococcioni, M and Colonna, N and Carnimeo, I and Dal Corso, A and De Gironcoli, S and Delugas, P and DiStasio, R A and Ferretti, A and Floris, A and Fratesi, G and Fugallo, G and Gebauer, R and Gerstmann, U and Giustino, F and Gorni, T and Jia, J and Kawamura, M and Ko, H-Y and Kokalj, A and K\"{u}\c{c}\"{u}kbenli, E and Lazzeri, M and Marsili, M and Marzari, N and Mauri, F and Nguyen, N L and Nguyen, H-V and Otero-de-la-Roza, A and Paulatto, L and Ponc\'{e}, S and Rocca, D and Sabatini, R and Santra, B and Schlipf, M and Seitsonen, A P and Smogunov, A and Timrov, I and Thonhauser, T and Umari, P and Vast, N and Wu, X and Baroni, S},
	month = nov,
	year = {2017},
	pages = {465901},
}

@article{Lee_UV_2020PRR,
	title = {First-principles approach with a pseudohybrid density functional for extended {Hubbard} interactions},
	volume = {2},
	number = {4},
	urldate = {2025-02-06},
	journal = {Phys. Rev. Res.},
	author = {Lee, Sang-Hoon and Son, Young-Woo},
	month = dec,
	year = {2020},
	pages = {043410},
}

@article{Pizzi_wannier90_2020,
	year = 2020,
	month = {jan},
	publisher = {{IOP} Publishing},
	volume = {32},
	number = {16},
	pages = {165902},
	author = {Giovanni Pizzi and Valerio Vitale and Ryotaro Arita and Stefan Bl{\"u}gel and Frank Freimuth and Guillaume G{\'{e}}ranton and Marco Gibertini and Dominik Gresch and Charles Johnson and Takashi Koretsune and Julen Iba{\~{n}}ez-Azpiroz and Hyungjun Lee and Jae-Mo Lihm and Daniel Marchand and Antimo Marrazzo and Yuriy Mokrousov and Jamal I Mustafa and Yoshiro Nohara and Yusuke Nomura and Lorenzo Paulatto and Samuel Ponc{\'{e}} and Thomas Ponweiser and Junfeng Qiao and Florian Thöle and Stepan S Tsirkin and Ma{\l}gorzata Wierzbowska and Nicola Marzari and David Vanderbilt and Ivo Souza and Arash A Mostofi and Jonathan R Yates},
	title = {Wannier90 as a community code: new features and applications},
	journal = {J. Phys.: Condens. Matter}
}

@book{lehoucq_arpack_1998,
	address = {Philadelphia},
	title = {{ARPACK} users' guide: solution of large-scale eigenvalue problems with implicitly restarted {Arnoldi} methods},
	shorttitle = {{ARPACK} users' guide},
	publisher = {Society for Industrial and Applied Mathematics},
	author = {Lehoucq, Richard B. and Sorensen, Danny C. and Yang, C. },
	year = {1998},
}

@misc{opencollab_arpack-ng,
	title = {Arpack-ng},
	author = {Opencollab},
    note ={[Accessed: 24-July-2025]},
     howpublished = {\url{https://github.com/opencollab/arpack-ng}}
}

@incollection{petsc_1997,
	address = {Boston, MA},
	title = {Efficient {Management} of {Parallelism} in {Object}-{Oriented} {Numerical} {Software} {Libraries}},
	isbn = {9781461219866},
	urldate = {2025-02-08},
	booktitle = {Modern {Software} {Tools} for {Scientific} {Computing}},
	publisher = {Birkh\"{a}user},
	author = {Balay, Satish and Gropp, William D. and McInnes, Lois Curfman and Smith, Barry F.},
	editor = {Arge, Erlend and Bruaset, Are Magnus and Langtangen, Hans Petter},
	year = {1997},
	pages = {163--202}
}

@article{Park2023NatComm,
	title = {Condensation of preformed charge density waves in kagome metals},
	volume = {14},
	number = {1},
	urldate = {2025-02-06},
	journal = {Nat. Commun.},
	author = {Park, Changwon and Son, Young-Woo},
	month = nov,
	year = {2023},
	pages = {7309},
}

@article{Kolmogorov2005PRB,
	title = {Registry-dependent interlayer potential for graphitic systems},
	volume = {71},
	number = {23},
	urldate = {2025-02-07},
	journal = {Phys. Rev. B},
	author = {Kolmogorov, Aleksey N. and Crespi, Vincent H.},
	month = jun,
	year = {2005},
	pages = {235415},
}

@article{Naik2019,
author = {Naik, Mit H. and Maity, Indrajit and Maiti, Prabal K. and Jain, Manish},
title = {Kolmogorov–Crespi Potential For Multilayer Transition-Metal Dichalcogenides: Capturing Structural Transformations in Moir\'{e} Superlattices},
journal = {J. Phys. Chem. C},
volume = {123},
number = {15},
pages = {9770-9778},
year = {2019}

}

@article{Peng2016,
  title = {Versatile van der Waals Density Functional Based on a Meta-Generalized Gradient Approximation},
  author = {Peng, Haowei and Yang, Zeng-Hui and Perdew, John P. and Sun, Jianwei},
  journal = {Phys. Rev. X},
  volume = {6},
  issue = {4},
  pages = {041005},
  numpages = {15},
  year = {2016},
  month = {Oct}
}

@Article{Acharya2025Nature,
author={Acharya, Rajeev
and Abanin, Dmitry A.
and Aghababaie-Beni, Laleh
and Aleiner, Igor 
and others
and AI, Google Quantum
and {Collaborators}},
title={Quantum error correction below the surface code threshold},
journal={Nature},
year={2025},
month={Feb},
day={01},
volume={638},
number={8052},
pages={920-926}
}

@article{Pezzagna2021APR,
    author = {Pezzagna, Sébastien and Meijer, Jan},
    title = {Quantum computer based on color centers in diamond},
    journal = {Appl. Phys. Rev.},
    volume = {8},
    number = {1},
    pages = {011308},
    year = {2021},
    month = {02}}

@article{Perdew1996,
  title = {Generalized Gradient Approximation Made Simple},
  author = {Perdew, John P. and Burke, Kieron and Ernzerhof, Matthias},
  journal = {Phys. Rev. Lett.},
  volume = {77},
  issue = {18},
  pages = {3865--3868},
  numpages = {0},
  year = {1996},
  month = {Oct}
}

@article{Baroni2001,
  title = {Phonons and related crystal properties from density-functional perturbation theory},
  author = {Baroni, Stefano and de Gironcoli, Stefano and Dal Corso, Andrea and Giannozzi, Paolo},
  journal = {Rev. Mod. Phys.},
  volume = {73},
  issue = {2},
  pages = {515--562},
  numpages = {0},
  year = {2001},
  month = {Jul}
}

@article{cortes2023,
title = {Ferroelectric response to interlayer shifting and rotations in trilayer hexagonal Boron Nitride},
journal = {J. Phys. Chem. Solids},
volume = {173},
pages = {111086},
year = {2023},
author = {Emilio A. Cort\'{e}s and Juan M. Florez and Eric Su\'{a}rez Morell}
}

@article{Haga2021,
  title = {Electronic states and modulation doping of hexagonal boron nitride trilayers},
  author = {Haga, Taishi and Matsuura, Yuuto and Fujimoto, Yoshitaka and Saito, Susumu},
  journal = {Phys. Rev. Mater.},
  volume = {5},
  issue = {9},
  pages = {094003},
  numpages = {11},
  year = {2021},
  month = {Sep},
  publisher = {American Physical Society}
}

@article{Yasuda2024,
author = {Kenji Yasuda  and Evan Zalys-Geller  and Xirui Wang  and Daniel Bennett  and Suraj S. Cheema  and Kenji Watanabe  and Takashi Taniguchi  and Efthimios Kaxiras  and Pablo Jarillo-Herrero  and Raymond Ashoori },
title = {Ultrafast high-endurance memory based on sliding ferroelectrics},
journal = {Science},
volume = {385},
number = {6704},
pages = {53-56},
year = {2024},
doi = {10.1126/science.adp3575},
URL = {https://www.science.org/doi/abs/10.1126/science.adp3575}
}

@article{Bian2024,
author = {Renji Bian  and Ri He  and Er Pan  and Zefen Li  and Guiming Cao  and Peng Meng  and Jiangang Chen  and Qing Liu  and Zhicheng Zhong  and Wenwu Li  and Fucai Liu },
title = {Developing fatigue-resistant ferroelectrics using interlayer sliding switching},
journal = {Science},
volume = {385},
number = {6704},
pages = {57-62},
year = {2024},
doi = {10.1126/science.ado1744},
URL = {https://www.science.org/doi/abs/10.1126/science.ado1744}
}

@article{kapfer2023science,
	title = {Programming twist angle and strain profiles in {2D} materials},
	volume = {381},
	issn = {0036-8075, 1095-9203},
	url = {https://www.science.org/doi/10.1126/science.ade9995},
	doi = {10.1126/science.ade9995},
	pages = {677--681},
	number = {6658},
	journaltitle = {Science},
	shortjournal = {Science},
	journal = {Science},
	author = {Kapfer, Ma\"{e}lle and Jessen, Bjarke S. and Eisele, Megan E. and Fu, Matthew and Danielsen, Dorte R. and Darlington, Thomas P. and Moore, Samuel L. and Finney, Nathan R. and Marchese, Ariane and Hsieh, Valerie and Majchrzak, Paulina and Jiang, Zhihao and Biswas, Deepnarayan and Dudin, Pavel and Avila, Jos\'{e} and Watanabe, Kenji and Taniguchi, Takashi and Ulstrup, S{\o}ren and B{\o}ggild, Peter and Schuck, P. J. and Basov, Dmitri N. and Hone, James and Dean, Cory R.},
	urldate = {2025-12-13},
	date = {2023-08-11},
	year = {2023},
	langid = {english},
}

@article{wu2023interface,
	title = {Enhanced Homogeneity of Moir\'{e} Superlattices in Double-Bilayer {WSe}$_{2}$ Homostructure},
	volume = {15},
	rights = {https://doi.org/10.15223/policy-029},
	issn = {1944-8244, 1944-8252},
	url = {https://pubs.acs.org/doi/10.1021/acsami.3c06949},
	doi = {10.1021/acsami.3c06949},
	pages = {48475--48484},
	number = {41},
	journaltitle = {{ACS} Applied Materials \& Interfaces},
	shortjournal = {{ACS} Appl. Mater. Interfaces},
	journal = {{ACS} Appl. Mater. Interfaces},
	author = {Wu, Biao and Zheng, Haihong and Li, Shaofei and Wang, Chang-Tian and Ding, Junnan and He, Jun and Liu, Zongwen and Wang, Jian-Tao and Liu, Yanping},
	urldate = {2025-12-13},
	date = {2023-10-18},
	year = {2023},
	langid = {english},
}
\end{document}